\documentclass[letterpaper]{article} 
\usepackage{aaai24}  
\usepackage{times}  
\usepackage{helvet}  
\usepackage{courier}  
\usepackage[hyphens]{url}  
\usepackage{graphicx} 
\usepackage{natbib}  
\usepackage{caption} 
\usepackage{newfloat}
\usepackage{listings}

\usepackage{algorithm}
\usepackage[utf8]{inputenc} 
\usepackage[T1]{fontenc}    
\usepackage{url}            
\usepackage{booktabs}       
\usepackage{amsfonts}       
\usepackage{nicefrac}       
\usepackage{microtype}      
\usepackage{xcolor}         
\usepackage{subfigure}
\usepackage{float}

\usepackage{graphicx}
\usepackage{mathtools}
\usepackage{amsmath}
\usepackage{enumitem}
\usepackage{amsthm}
\DeclareMathAlphabet{\mathcal}{OMS}{cmsy}{m}{n}

\usepackage{amssymb}
\usepackage{multirow}
\usepackage{booktabs}
\usepackage{svg}

\DeclareCaptionStyle{ruled}{labelfont=normalfont,labelsep=colon,strut=off} 
\floatstyle{ruled}
\newfloat{listing}{tb}{lst}{}
\floatname{listing}{Listing}
\title{A Generalized Neural Diffusion Framework on Graphs}
\author{
    Yibo Li\textsuperscript{\rm 1},
    Xiao Wang\textsuperscript{\rm 2},
    Hongrui Liu\textsuperscript{\rm 3},
    Chuan Shi\textsuperscript{\rm 1}\thanks{Corresponding author.}
}
\affiliations{
    \textsuperscript{\rm 1}Beijing University of Posts and Telecommunications\\
    \textsuperscript{\rm 2}Beihang University\\
    \textsuperscript{\rm 3}Ant Group\\

    \{yiboL, shichuan\}@bupt.edu.cn, xiao\_wang@buaa.edu.cn, liuhongrui.lhr@antgroup.com
}

\usepackage{bibentry}

\begin{document}

\maketitle

\begin{abstract}
Recent studies reveal the connection between GNNs and the diffusion process, which motivates many diffusion-based GNNs to be proposed. However, since these two mechanisms are closely related, one fundamental question naturally arises: \textit{Is there a general diffusion framework that can formally unify these GNNs?} The answer to this question can not only deepen our understanding of the learning process of GNNs, but also may open a new door to design a broad new class of GNNs. In this paper, we propose a general diffusion equation framework with the fidelity term, which formally establishes the relationship between the diffusion process with more GNNs. Meanwhile, with this framework, we identify one characteristic of graph diffusion networks, i.e., the current neural diffusion process only corresponds to the first-order diffusion equation. However, by an experimental investigation, we show that the labels of high-order neighbors actually exhibit monophily property, which induces the similarity based on labels among high-order neighbors without requiring the similarity among first-order neighbors. This discovery motives to design a new high-order neighbor-aware diffusion equation, and derive a new type of graph diffusion network (HiD-Net) based on the framework. With the high-order diffusion equation, HiD-Net is more robust against attacks and works on both homophily and heterophily graphs. We not only theoretically analyze the relation between HiD-Net with high-order random walk, but also provide a theoretical convergence guarantee. Extensive experimental results well demonstrate the effectiveness of HiD-Net over state-of-the-art graph diffusion networks.
\end{abstract}

\section{Introduction}
\label{section:introduction}
Graphs, such as traffic networks,  social networks, citation networks, and molecular networks, are ubiquitous in the real world. Recently, Graph Neural Networks (GNNs), which are able to effectively learn the node representations based on the message-passing manner,  have shown great popularity in tackling graph analytics problems. So far, GNNs have significantly promoted the development of graph analysis towards real-world applications. e.g, node classifification \citep{nodeclassification1, EvenNet, wang}, link prediction \citep{linkprediction1, linkprediction2}, graph reconstruction \citep{l2g2g}, subgraph isomorphism counting \citep{xt}, and graph classifification \citep{graphclassification1, graphclassification2}.

Some recent studies show that GNNs are in fact intimately connected to diffusion equations  \citep{grand, dgc, grand++}, which can be considered as information diffusion on graphs.  Diffusion equation interprets GNNs from a continuous perspective \citep{grand} and provides new insights to understand existing GNN architectures, which motives some diffusion-based GNNs. For instance, \citep{dgc} proposes continuous graph diffusion.  \citep{grand++} utilizes the diffusion process to handle oversmoothing issue. \citep{robust} considers a graph as a discretization of a Riemannian manifold and studies the robustness of the information propagation process on graphs. Diffusion equation can also build a bridge between traditional GNNs and control theory \citep{control}. Although the diffusion process and graph convolution are closely related, little effort has been made to answer: \textit{Is there a unified diffusion equation framework to formally unify the current GNN architectures?} A well-informed answer can deepen our understanding of the learning mechanism of GNNs, and may inspire to design a broad new class of GNNs based on diffusion equation. 

Actually, \citep{grand} has explained GCN \citep{gcn} and GAT \citep{gat} from diffusion equation. However, with more proposed GNN architectures, it is highly desired to formally revisit the relation between diffusion equation and GNNs. In this paper, we discover that many GNN architectures substantially can be unified with a general diffusion equation with the fidelity term, such as GCN/SGC [22], APPNP[14], GAT [19], AMP [16], DAGNN [15]. Basically, the diffusion equation describes that the change of a node representation depends on the movement of information on graphs from one node to its neighbors, and the fidelity term constraints that the change of a node representation depends on the difference with its initial feature. Furthermore, we show that the unified diffusion framework can also be derived from an energy function, which explains the whole framework as an energy minimization process in a global view. Compared with other unified frameworks \citep{zhu, ma}, our framework is from the diffusion perspective, which has many advantages. For example, diffusion-based methods are able to address the common plights of graph learning models such as oversmoothing \citep{grand, dgc}. What's more, the diffusion equation can be seen as partial differential equations (PDEs) \citep{grand}, thus introducing many schemes to solve the graph diffusion equation such as explicit scheme, implicit scheme, and multi-step scheme, some of which are more stable and converge faster. 

Based on the above findings, we can see that the diffusion process employed by most current GNNs just considers the first-order diffusion equation, which only diffuses messages among 1-hop neighbors. That is, the first-order diffusion has the underlying homophily assumption among 1-hop neighbors. While we empirically discover that the labels of 2-hop neighborhoods actually appear monophily property \citep{monophily}, i.e., nodes may have extreme preferences for a particular attribute which are unrelated to their own attribute and 1-hop neighbors' attribute, but are more likely to be similar with the attribute of their 2-hop neighbors. Simply put, monophily can induce a similarity among 2-hop neighbors without requiring similarity among 1-hop neighbors. So when the 1-hop neighbors are heterophily-dominant or have noise, the 2-hop neighbors will provide more relevant context.
Therefore, a more practical diffusion process should take both the first-order and second-order neighbors into account. 
\textit{How can we design a new type of graph diffusion networks satisfying the above requirement based on our framework?}

In this paper, we design a new high-order neighbor-aware diffusion equation in our proposed diffusion framework, and then derive a High-order Graph Diffusion Network (HiD-Net). Specifically, our model simultaneously combines the first-order and second-order diffusion process, then we regularize the diffusion equation by minimizing the discrepancy between the estimated and the original graph features. The whole diffusion equation is finally integrated into the APPNP architecture. With second-order diffusion equation, HiD-Net is more robust against the attacks and more general on both homophily and heterophily graphs. We theoretically prove that HiD-Net is essentially related with the second-order random walk. We also provide the convergence guarantee that 
HiD-Net will converge to this random walk’s limit distribution as the number of layers increases, and meanwhile, the learned representations do not converge to the same vector over all nodes.
The contributions of this paper are summarized as follows:

\begin{itemize}
\item We propose a novel generalized diffusion graph framework, consisting of diffusion equation and fidelity term. This framework, formally establishing the relation between diffusion process with a wide variety of GNNs, describes a broad new class of GNNs based on the discretized diffusion equations on graphs and provides new insight to the current graph diffusion/neural networks.
  
\item We discover the monophily property of labels, and based on our diffusion framework, we propose a high-order graph diffusion network, HiD-Net, which is more general and robust. We theoretically build the relation between HiD-Net and second-order random walk, together with the convergence guarantee.

\item  Our extensive experiments on both the homophily and heterophily graphs clearly show that HiD-Net outperforms the popular GNNs based on diffusion equation. 
\end{itemize}

\section{Related Work}
\label{section:relatedwork}
\label{gen_inst}

\textbf{Graph convolutional networks.}
Recently, graph convolutional network (GCN) models  \citep{spectral, chebnet, gcn, gat, graphsage} have been widely studied.
Based on the spectrum of graph Laplacian,     \citep{spectral} generalizes CNNs to
graph signal.
Then \citep{chebnet} further improves the efficiency by employing the Chebyshev expansion of the graph Laplacian.
\citep{gcn} proposes to only aggregate the node features from the one-hop neighbors and simplifies the convolution operation.
\citep{gat} introduces the attention mechanisms to learn aggregation weights adaptively. 
\citep{graphsage} uses various ways of pooling for aggregation.
More works on GNNs can be found in surveys \citep{survey1, survey2}.

\textbf{Diffusion equation on graphs.}
Graph Heat Equation (GHE) \citep{ghe}, which is a well-known generalization of the diffusion equation on graph data, models graph dynamics with applications in spectral graph theory. 
GRAND \citep{grand} studies the discretized diffusion PDE on graphs and applies different numerical schemes for their solution.
GRAND++ \citep{grand++} mitigates the oversmoothing issue of graph neural networks by adding a source term.
DGC \citep{dgc} decouples the terminal time and propagation steps of linear GCNs from a perspective of graph diffusion equation, and analyzes why linear GCNs fail to benefit from deep layers.
ADC \citep{adc} strategies to automatically learn the optimal diffusion time from the data.
However, these works focus on specific graph diffusion network, thus there is not a framework to formally unify the GNNs.

\textbf{The unified GNN framework.} \citep{zhu} establishes a connection between different propagation mechanisms with a unified optimization problem, and finds out that the proposed propagation mechanisms are the optimal solution for optimizing a feature-fitting function over a wide class of graph kernels with a graph regularization term. \citep{ma} establishes the connections between the introduced GNN models and a graph signal denoising problem with Laplacian regularization. It essentially is still an optimization solving framework from the perspective of signal denoising. However, our framework is 
based on the diffusion equation, where the advantages are two fold: one is that diffusion-based methods are able to address the oversmoothing problem \citep{grand, dgc}. The other is that the diffusion equation can be seen as partial differential equations (PDEs) \citep{grand} and thus can introduce many schemes that have many good properties such as fast convergence rate and high stability.

\vspace{-3mm}
\section{The Generalized Diffusion Graph Framework}
\label{section:framework}

\textbf{Notations.} Consider an undirected graph as $ G = (V, E)$ with adjacency matrix  $\mathbf{A}\in \mathbb{R}^{n \times n}$, where $V$ contains $n$ nodes $\{v_1, \dots, v_n\}$ and $E$ is the set of edges. The initial node feature matrix is denoted as $\mathbf{X}^{(0)}\in \mathbb{R}^{n \times q}$, where $q$ is the dimension of node feature. 
 We denote the neighbors of node $i$ at
exactly $k$ hops/steps away as $N_k(i)$.  For example, $N_1(i) = \{j : (i, j) \in E\} $ are the immediate neighbors of $i$.


 Diffusion is a physical process that equilibrates concentration differences without creating or destroying mass. This physical observation can be easily cast in the diffusion equation, which is a parabolic partial differential equation.
Fick’s law of diffusion describes the equilibration property \citep{anisotropicdiffusion}:
\vspace{-0mm}
\begin{equation}
    J=-f\cdot\nabla u,
    \label{eq:fick}
\end{equation}

where $J$ is the diffusion flux, which measures the amount of substance that flows through a unit area during a unit time interval. $f$ is the diffusivity coefficient, which can be constant or depend on time and position. $\nabla u$ is the concentration gradient.
This equation states that a concentration gradient $\nabla u$ causes a flux $J$ which aims to compensate for this gradient. The observation that a change in concentration in any part of the system is due to the influx and outflux of substance into and out of that part of the system can be expressed by the continuity equation: 

\vspace{-1mm}
\begin{equation}
    \dfrac{\partial u}{\partial t}=-\operatorname{div} J,
\end{equation}

where $t$ denotes the time.
Plugging in Fick’s law ~\eqref{eq:fick} into the continuity equation we end up with the diffusion equation:

\vspace{-1mm}
\begin{equation}
    \dfrac{\partial u}{\partial t} = \operatorname{div} (f \cdot \nabla u).
\end{equation}

As $\operatorname{div}$ is the sum of the second derivatives in all directions, please note that normal first order derivatives and second order derivatives are on continuous space and can not be generalized directly to graph which is on discrete space. As \citep{grand} defined, the first derivative is the difference between the feature of a node and its neighbor. And the second derivative can be considered as the difference between the first derivatives of the node itself and its neighbors. 

\begin{figure}[H] 
\centering 
\includegraphics[width=0.35\textwidth]{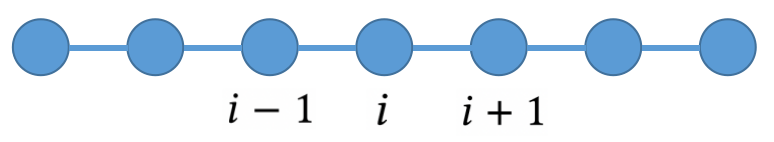} 
\caption{Chain graph} 
\label{fig:chain} 
\end{figure}

For better illustration, we provide an example. Consider a chain graph in Figure ~\ref{fig:chain}, where $i$, $i+1$, and $i-1$ are the indexes of the nodes. The feature of node $i$ is denoted as $x_i$.

\vspace{-3mm}

\vspace{-5mm}

The first order derivatives on node $i$ is defined as $x_{i+1}-x_{i}$ and $x_{i} - x_{i-1}$. The diffusion flux from node $j$ to node $i$ at time $t$ on a graph is:

\vspace{-1mm}
\begin{equation}
    J_{ij}^{(t)}=-f\cdot (\nabla x)_{ij}^{(t)}=-f(x_j^{(t)}-x_i^{(t)}).
\end{equation}

The second order derivative is the difference of first order derivatives: $(x_{i+1}-x_{i})-(x_{i}-x_{i-1})=(x_{i+1}-x_{i})+(x_{i-1} - x_{i})$. We notice that the chain graph only has one dimension, so the divergence of node $i$ on a chain graph is equal to its second order derivative: $\operatorname{div}(\nabla x_{i}) = (x_{i+1}-x_{i})+(x_{i-1} - x_{i})$. 

Thus, on a normal graph, we have the generalized form: $\operatorname{div}(\nabla x_{i}) = \sum_{j \in N_1(i)} (x_j-x_i)$.

Here we normalize the diffusion process utilizing the degree of the nodes to down-weight the high-degree neighbors, and we have $\operatorname{div}(\nabla x_{i}) = \sum_{j \in N_1(i)} \dfrac{\tilde{A}_{ij}}{\sqrt{\tilde{d_i}}\sqrt{\tilde{d_j}}}(x_j-x_i)$, where $\tilde{A}_{ij}$ is the element of $\tilde{\mathbf{A}} = \mathbf{A} + \mathbf{I}$, and $d_i = \sum_{j}\tilde{A}_{ij}$.


So the diffusion equation on node $i$ can be defined as:

\vspace{-2mm}
\begin{equation}
\begin{aligned} 
    \frac{\partial x_i^{(t)}}{\partial t}&=-\operatorname{div}J_{ij}^{(t)}
    =\operatorname{div}[f (\nabla x)_{ij}^{(t)}]\\
    &=f\sum_{j \in N_1(i)} \dfrac{\tilde{A}_{ij}}{\sqrt{\tilde{d_i}}\sqrt{\tilde{d_j}}}(x_j^{(t)}-x_i^{(t)}).
\end{aligned}
\label{eq:div}
\end{equation}

The diffusion equation models the change of representation $x_i^{(t)}$ with respect to $t$, which depends on the difference between the nearby nodes, implying that the greater the difference between a node and its neighbors, the faster it changes.

However, how fast $x_i^{(t)}$ changes should not only depend on the representation difference between node $i$ and its neighbors, otherwise, it will cause oversmoothing issue, i.e., as the diffusion process goes by, the nodes are not distinguishable. 
Based on this phenomenon, we think that the representation change of $x_i^{(t)}$ should be also related with the node feature $x_i^{(0)}$ itself, i.e., if the difference between $x_i^{(t)}$ and $x_i^{(0)}$ is small, the change of $x_i^{(t)}$ should also be small. Then we add another fidelity term and obtain our general graph diffusion framework as follows:

\vspace{-2mm}
\begin{equation}
    \dfrac{\partial x_i^{(t)}}{\partial t} = \alpha (x_i^{(0)} - x_i^{(t)}) + \beta \operatorname{div}(f (\nabla x)_{ij}^{(t)}),
     \label{eq:diffusion}
\end{equation}
where $\alpha$, $\beta$ are coefficients.

\textbf{Remark 1.}
~\eqref{eq:diffusion} can be derived from the energy function:

\vspace{-4mm}
\begin{equation}
E(x):=\int_{\Omega}\left(\alpha \cdot(x_i-x_i^{(0)})^{2}+\beta \cdot|f (\nabla x)_{ij}|^{2} \right)d\theta,
\label{eq:energyfunction}
\end{equation}
where $\theta$ represents the position of the nodes, and $\Omega$ represents the entire graph domain. The corresponding Euler–Lagrange equation, which gives the necessary condition for an extremum of ~\eqref{eq:energyfunction},  is given by:

\vspace{-2mm}
\begin{equation}
    0=\alpha (x_i - x_i^{(0)}) + \beta \operatorname{div}(f (\nabla x)_{ij}).
    \label{eq:euler}
\end{equation}
~\eqref{eq:euler} can also be regarded as the steady-state equation of ~\eqref{eq:diffusion}. Based on the energy function, we can see that ~\eqref{eq:diffusion} constrains space variation and time variation of the diffusion process, indicating that the representations of the graph nodes will not change too much between nearby nodes, as well as not change too much from the initial features.

\textbf{Remark 2.} The framework~\eqref{eq:diffusion} is closely related to many GNNs, such as GCN/SGC \citep{sgc}, APPNP \citep{appnp}, GAT \citep{gat}, AMP \citep{amp}, DAGNN~\citep{dagnn}, as demonstrated by the following propositions. 
We provide the proofs of all the subsequent propositions in Appendix.

\textbf{Proposition 1.} With $\alpha=0$, $\beta=1$, $\Delta t = 1$ and $f=1$ in  ~\eqref{eq:diffusion}, the diffusion process in SGC/GCN  is:

\vspace{-2mm}
\begin{equation}
    \dfrac{\partial x_i^{(t)}}{\partial t}= \operatorname{div} ((\nabla x)_{ij}^{(t)}).
    \label{eq:sgc}
\end{equation}

\textbf{Proposition 2.} Introducing $\eta$ as coefficient, with $\alpha=1$, $\beta=1-\dfrac{1}{\eta}$, $\Delta t = 1$   and $f=1$ in  ~\eqref{eq:diffusion}, the diffusion process in APPNP  is:
\vspace{-1mm}
\begin{equation}
    \dfrac{\partial x_i^{(t)}}{\partial t}= (x_i^{(0)} - x_i^{(t)})+(1-\dfrac{1}{\eta} ) \operatorname{div} ((\nabla x)_{ij}^{(t)} ).
    \label{eq:appnp}
\end{equation}







\textbf{Proposition 3.} With $\alpha=0$, $\beta=1$, $\Delta t = 1$   and the learned similarity coefficient $f_{ij}^{(t)}$ between nodes $i$ and $j$ at time $t$ in  ~\eqref{eq:diffusion}, the diffusion process in GAT  is:

\vspace{-2mm}
\begin{equation}
    \dfrac{\partial x_i^{(t)}}{\partial t}= \operatorname{div} (f_{ij}^{(t)}(\nabla x)_{ij}^{(t)}).
    \label{eq:gat}
\end{equation}



    

\textbf{Proposition 4.} With stepsize $\epsilon$ and coefficient $\lambda$, $\beta_{i}^{(t)} =\max \left(1-\frac{\epsilon \lambda}{\left\|(1-2 \epsilon(1-\lambda)) \mathbf{X}_i^{(t)}+2 \epsilon(1-\lambda) \tilde{\mathbf{A}} \mathbf{X}_i^{(t)}-\left(\mathbf{X}^{(0)}\right)_{i}\right\|_{2}}, 0\right)$. Let $\alpha = 1-\beta_i^{(t)}$, $\beta = 2\epsilon(1-\lambda) \beta_i^{(t)}$, $\Delta t = 1$   and $f=1$ in ~\eqref{eq:diffusion}, the diffusion process in AMP  is:

\vspace{-3mm}
\begin{equation}
    \dfrac{\partial x_i^{(t)}}{\partial t} = (1-\beta_i^{(t)}) (x_i^{(0)} - x_i^{(t)}) + 2\epsilon(1-\lambda) \beta_i^{(t)} \operatorname{div}( (\nabla x)_{ij}^{(t)}).
    \label{eq:amp}
\end{equation}

\textbf{Proposition 5.} With $\alpha + \beta=1$, $\Delta t = 1$   and the learned coeffiecient $f(t)$ at time $t$ satisfying $ \sum_{t=0}^{T}(\beta f(t))^{t} =\frac{1}{\alpha}$ in ~\eqref{eq:diffusion}, the diffusion process in DAGNN is:  

\vspace{-4mm}
\begin{equation}
    \dfrac{\partial x_i^{(t)}}{\partial t}=\alpha (x_i^{(0)} - x_i^{(t)})+\beta  \operatorname{div} (f(t)((\nabla x)_{ij}^{(t)} )). 
    \label{eq:dagnn}
\end{equation}





\section{The High-order Graph Diffusion Network}
\label{section:model}
\subsection{High-order Graph Diffusion Equation}
\label{subsection:dmp}
\vspace{-4mm}
\begin{figure}[htbp]
\centering  
\subfigure[]{   
\setlength{\abovecaptionskip}{0.cm}
\begin{minipage}{.3\linewidth}
\centering    
\includegraphics[scale=0.26]{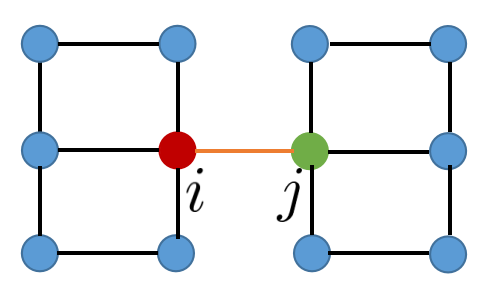}  
\label{fig:context1}
\end{minipage}
}
\subfigure[]{ 
\begin{minipage}{.3\linewidth}
\centering    
\includegraphics[scale=0.26]{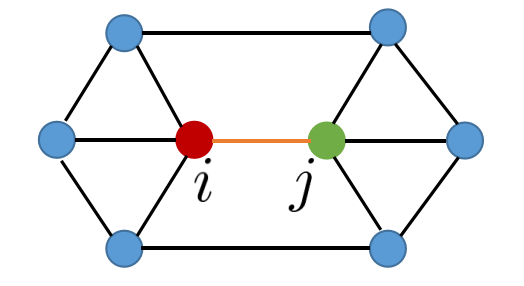}
\label{fig:context2}
\end{minipage}
}
\subfigure[]{ 
\begin{minipage}{.3\linewidth}
\centering    
\includegraphics[scale=0.26]{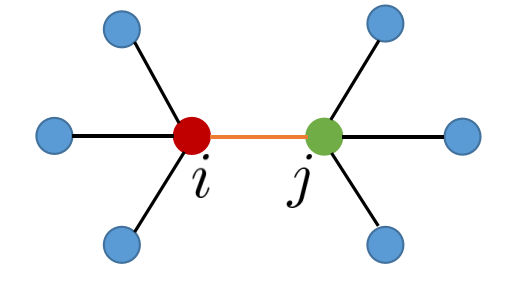}
\label{fig:context3}
\end{minipage}
}
\caption{Illustration of the same node pair in different contexts.}
    \label{fig:context}
\end{figure}
\vspace{-4mm}
In the first-order diffusion process, the diffusion equation only considers 1-hop neighbors. As shown in Figure ~\ref{fig:context}, the nodes in (a), (b), and (c) are the same, but the structures are different. The first-order diffusion flux from node $j$ to node $i$ will be the same, even if the local structure of node $i$ and $j$ is very different. Based on the specific local environments, the diffusion flux should be either different, so as to provide more additional information and make the learned representations more discriminative. To better understand the effect of local structures, we conduct an experiment on six widely used graphs to evaluate the effect of 2-hop neighbors. First, we have the following definition of $k$-hop neighbor similarity score.

\textbf{Definition 1.} Let $y_i$ be the label of node $i$, the $k$-hop neighbor similarity score $h_k=\dfrac{|\sum _{i \in V} \mathbf{1}_{y_i = \mathbf{O}(\{y_{j}, j\in N_k(i)\})}|}{|N_k(i)|}$, and $h_{a+b}=\dfrac{|\sum _{i \in V} \mathbf{1}_{y_i = \mathbf{O}(\{y_{j}, j\in \{N_a(i),N_b(i)\}\})}|}{|\{N_a(i),N_b(i)\}|}$, where $\mathbf{O}(\{y_{j}, j\in N_k(i)\})$ represents the element with the highest frequency in $\{y_{j}, j\in N_k(i)\}$.

The similarity score is based on node labels, and higher similarity score implies the labels of a node and its $k$-hop neighbors are more consistent. The scores of the six graphs are shown in Table ~\ref{table:hk}.  Interestingly, we find that the labels of 2-hop neighbors show monophily property \citep{monophily}, i.e., as can be seen from both the homophily graphs (Cora, Citeseer, Pubmed) and heterophily graphs (Chameleon, Squirrel, Actor), without requiring the similarity among first-order
neighbors, the second-order neighbors are more likely to have the same labels. 

\begin{table}[htpb]
\fontsize{9pt}{\baselineskip}\selectfont
\Large

 \centering
 \resizebox{\linewidth}{!}{
 \begin{tabular}{p{0.5cm}p{1cm}p{1cm}p{1cm}p{1.5cm}p{1.2cm}p{0.9cm}}
\hline

& cora     & citeseer & pubmed & chameleon & squirrel & actor \\
\hline
$h_1$ & 0.8634  & 0.7385 &  0.7920 & 0.2530 & 0.1459 &  0.2287\\
$h_2$ & 0.8696 & 0.8476 & 0.7885 & 0.3131 & 0.1600 & 0.3716\\
$h_{1+2}$ & 0.8737 & 0.8206 & 0.7880 & 0.3070 & 0.1530 & 0.3363 \\
\hline

 \end{tabular}
 }\caption{The similarity scores of six graphs.}
 \label{table:hk}
\end{table}

To take advantage of 2-hop neighbors, we regularize the gradient $\nabla x_i$ utilizing the average gradient of 1-hop neighbors: 

\vspace{-5mm}
\begin{equation}
    \overline{(\nabla x)_j} = \operatorname{avg}(\nabla x_{jk})=\sum_{k\in N_1(j)}  \dfrac{\tilde{A}_{jk}}{\sqrt{\tilde{d_j}} \sqrt{\tilde{d_k}}}(x_k - x_j).
\label{eq:G}
\end{equation}

\vspace{-2mm}
We propose the high-order graph diffusion equation:
\vspace{-3mm}
\begin{equation}
    \dfrac{\partial x_i^{(t)}}{\partial t}=\alpha (x_i^{(0)} - x_{i}^{(t)})+\beta  \operatorname{div} (f(\nabla x_{ij}^{(t)}) +\gamma \overline{(
    \nabla x)_j^{(t)}}),
    \label{eq:ourmodel}
\end{equation}
where $\gamma$ is the parameter of the regularization term. The iteration step of ~\eqref{eq:ourmodel} is:

\vspace{-3mm}
\begin{equation}
\begin{split}
x_i^{t+\Delta t}&=\alpha \Delta t x_i^{(0)} + (1-\alpha \Delta t ) x_i^{(t)}\\&+\beta \Delta t  \operatorname{div} (f(\nabla x_i^{(t)} )) + \beta\gamma \Delta t \operatorname{div} (\overline{(
    \nabla x)_j^{(t)}}),
\label{eq:model}
\end{split}
\end{equation}
which is the diffusion-based message passing scheme (DMP) of our model. We can see that DMP utilizes the 2-hop neighbors' information, where the advantages are two-fold: one is that the 2-hop neighbors capture the local environment around a node, 
even if there are some abnormal features among 1-hop neighbors, their negative effect can still be alleviated by considering a larger neighborhood size, making the learning process more robust.
 The other is that the monophily property of 2-hop neighbors provides additional stronger correlation with labels, thus even if the 1-hop neighbors may be heterophily, DMP can still make better predictions with information diffused from 2-hop neighbors. 

\textbf{Comparison with other GNN models.} Though existing GNN iteration steps can capture high-order connectivity through iterative adjacent message passing, they still have their limitations while having the same time complexity as DMP. DMP is superior because it can utilize the monophily property, adjust the balance between first-order and second-order neighbors, and is based on diffusion equation which has some unique characteristics. More comparisons are discussed in Appendix.

\subsection{Theoretical Analysis}
Next, we theoretically analyze some properties of our diffusion-based message passing scheme.

\textbf{Definition 2.} Consider a surfer walks from node $j$ to $i$ with probability $P_{ij}$. Let $\mathbb{X}_t$ be a random
variable representing the node visited by the surfer at time $t$. The probability $P_{ij}$ can be represented as a conditional probability $\mathbb{P}\left[\mathbb{X}_{t}=i \mid \mathbb{X}_{t-\Delta t}=j\right]$. Let

\vspace{-2mm}
\begin{equation}
    P_{ij}=\left\{  
    \begin{array}{lr}
    1-(\alpha + \beta) \Delta t, &  i=j\\  
    (\beta - \beta \gamma) \Delta t \hat{\tilde{A}}_{ij}, & j \in N_1(i)\\  
    \beta \gamma \Delta t B_{ij}, & j \in N_2(i)\\
    \alpha \Delta t, & restart,\\
    \end{array}  
    \right.
    \label{eq:p}
\end{equation}
\vspace{-1mm}
where $\hat{\tilde{\mathbf{A}}} = \tilde{\mathbf{D}}^{-\frac{1}{2}}\tilde{\mathbf{A}}\tilde{\mathbf{D}}^{-\frac{1}{2}}$, $B_{ij}$ is the element of $\mathbf{B}=\hat{\tilde{\mathbf{A}}}^2 $, and restart means that the node $i$ will teleport back to the initial root node $i$. Based on the definition, we have the following propositions.

\textbf{Proposition 6.} Given the probability $\mathbb{H}^{(t)}_{ij} = \mathbb{P}\left[\mathbb{X}_{t}=i \mid \mathbb{X}_{0}=j\right]$, DMP~\eqref{eq:model} is equivalent to the second-order random walk with the transition probability $P_{ij}$ in ~\eqref{eq:p}:

\vspace{-3mm}
\begin{equation}
    x_i^{(t)} = \sum_{j \in V} \mathbb{H}^{(t)}_{ij} x_j^{(0)}.
\end{equation}
\vspace{-2mm}

\textbf{Proposition 7.} With $f=1$, $\alpha, \beta, \gamma, \Delta t \in \left(0, 1\right]$, DMP~\eqref{eq:model} converges, i.e., when $t\rightarrow \infty$, $\mathbf{X}^{(\infty)} = \alpha((\alpha + \beta) \mathbf{I} - \beta(1- \gamma)  \hat{\tilde{\mathbf{A}}} -\beta \gamma \hat{\tilde{\mathbf{A}}}^2)^{-1} \mathbf{X}^{(0)}$.

\textbf{Proposition 8.} When $t\rightarrow \infty$, the representations of any two nodes on the graph will not be the same as long as the two nodes have different initial features, i.e., 
$\forall i,j\in V$, if $x_i^{(0)}\neq x_j^{(0)}$, then $x_i^{(t)} \neq x_j^{(t)}$ as $t\rightarrow \infty$.

The proofs of the above propositions are in Appendix.

    


\subsection{Our Proposed HiD-Net}

To incorporate the high-order graph diffusion DMP~\eqref{eq:model} into deep neural networks, we introduce High-Order Graph Diffusion Network (HiD-Net). In this work, we follow the decoupled way as proposed in APPNP \citep{appnp}:
\vspace{-2mm}
\begin{equation}
\mathbf{Y}'=\mathbf{DMP}\left(l_{\omega}\left(\mathbf{X}^{(0)}\right), t, \Delta t, \alpha, \beta, \gamma \right),
\end{equation}
where $l_{\omega}$ is a representation learning model such as an MLP network, $\omega$ is the learnable parameters in the model. The training objective is to minimize the cross entropy loss defined by the final prediction $\mathbf{Y}'$ and labels for training data. Because of DMP, HiD-Net is more robust and works well on both homophily and heterophily graphs in comparison with other graph diffusion networks. 

\textbf{Time complexity.} The time complexity of HiD-Net can be optimized as  $\mathcal{O}(n^2\zeta)$, which is the same as the propagation step of GCN, where $n$ is the number of the nodes, $\zeta$ is the dimension of the feature vector. 
We provide the proof in Appendix.




\section{Experiments}
\label{section:experiments}
\label{others}


\subsection{Node Classification}
\label{sec:nodeclassification}
\textbf{Datasets.} For comprehensive comparison, we use seven real-world datasets to evaluate the performance of node classification. They are three citation graphs, i.e., Cora, Citeseer, Pubmed \citep{gcn}, two Wikipedia networks, i.e., Chameleon and Squirrel \citep{hetero}, one Actor co-occurrence network Actor \citep{hetero}, one Open Graph Benchmark(OGB) graph ogbn-arxiv\citep{ogbn}. Among the seven datasets, Cora, Citeseer, Pubmed and ogbn-arxiv are homophily graphs, Chameleon, Squirrel, and Actor are heterophily graphs. Details of datasets are in Appendix.

    

\linespread{0.7}
\begin{table*}[htpb]
    \centering
    \fontsize{9pt}{\baselineskip}\selectfont

    \begin{tabular}{c|c|c|c|c|c|c|c|c|c}
    \toprule
        Datasets & Metric & GCN & GAT & APPNP & GRAND & GRAND++ & DGC & ADC & \textbf{HiD-Net}\\
    \midrule
        \multirow{3}*{Cora} 
        &F1-macro & 81.5$\pm0.6$ & 79.7$\pm0.4$ & 82.2 $\pm 0.5$ &79.4$\pm2.4$ & 81.3$\pm3.3$ & 82.1$\pm 0.1$ &80.0$\pm1.0$ &\textbf{82.8$\mathbf{\pm0.6}$}  \\
        &F1-micro &82.5 $\pm 0.6$ & 80.1$\pm0.8$ &83.2$\pm0.2$ & 80.1$\pm2.7$ & 82.95 $\pm 1.4$ &83.1$\pm0.1$ &81.0$\pm0.7$ & \textbf{84.0$\mathbf{\pm0.6}$}  \\
        &AUC & 97.3 $\pm 0.1$& 96.4 $\pm 0.5$ & 97.5$\pm0.1$ &96.0$\pm0.3$ & 97.3$\pm0.5$ & 97.2$\pm0.0$ &97.1$\pm0.1$ & \textbf{97.6$\mathbf{\pm0.0}$}  \\
    \midrule
        \multirow{3}*{Citeseer} 
        &F1-macro & 66.4 $\pm 0.4$& 68.5 $\pm0.3$ &67.7$\pm0.6$ &64.9$\pm1.5$ &66.4$\pm2.6$ &68.3$\pm0.4$ &47.0$\pm1.4$ & \textbf{69.5$\mathbf{\pm0.6}$}  \\
        &F1-micro &69.9 $\pm0.5$ & 72.2 $\pm0.3$ & 71.0 $\pm0.4$ &68.6$\pm1.7$ &70.9$\pm2.3$ &72.5$\pm0.4$ &53.7$\pm1.5$ &\textbf{73.2$\mathbf{\pm0.2}$}  \\
        &AUC & 89.9 $\pm 0.4$& 90.2 $\pm0.1$ &90.3 $\pm 0.0$ &89.5$\pm0.8$ & 91.2$\pm2.8$ &91.0 $\pm0.0$ &87.1$\pm1.1$ &\textbf{91.5$\mathbf{\pm0.1}$}  \\
    \midrule
        \multirow{3}*{Pubmed} 
        &F1-macro & 78.4 $\pm 0.2$& 76.7 $\pm0.5$ &79.3$\pm0.2$ &77.5$\pm3.2$ &78.9$\pm2.5$ &78.4$\pm0.1$ &73.7$\pm2.3$ &\textbf{80.1$\mathbf{\pm0.1}$}  \\
        &F1-micro &79.1 $\pm0.4$ & 77.3 $\pm0.4$ &79.9$\pm0.3$ & 78.0$\pm3.2$& 79.8$\pm1.6$ &79.2$\pm0.1$ &74.3$\pm2.3$ &\textbf{81.1$\mathbf{\pm0.1}$}  \\
        &AUC & 91.2 $\pm0.2$& 90.3$\pm0.5$ & \textbf{92.2$\mathbf{\pm0.1}$ } & 90.7$\pm1.6$& 91.5$\pm2.2$ &92.0$\pm0.0$ &89.1$\pm1.9$ &\textbf{92.2$\mathbf{\pm0.1}$ } \\
    \midrule
        \multirow{3}*{Chameleon} 
        &F1-macro & 38.5 $\pm2.1$& 45.0 $\pm1.0$ &57.5$\pm 1.0 $& 35.7$\pm1.8$ & 46.3$\pm2.4$ &58.0$\pm0.1$ &32.6$\pm0.6$ &\textbf{61.0$\mathbf{\pm0.3}$}  \\
        &F1-micro &41.8 $\pm1.2$ & 44.4 $\pm1.9$ &57.1$\pm1.4$  & 37.7$\pm1.5$ & 45.7$\pm3.4$ &58.2$\pm0.1$ &33.2$\pm0.5$ &\textbf{60.8$\mathbf{\pm0.7}$}  \\
        &AUC & 69.8$\pm0.5$& 75.5 $\pm1.0$ &85.0$\pm0.6$  &69.0 $\pm1.3$ &74.8 $\pm2.8$ &82.4$\pm0.0$ &63.7$\pm0.8$ &\textbf{85.2$\mathbf{\pm0.3}$ } \\
    \midrule
        \multirow{3}*{Squirrel} 
        &F1-macro & 25.2 $\pm 1.2$& 26.5 $\pm 1.3$&41.1$\pm1.1$ & 24.7$\pm2.0 $& 30.5$\pm3.7 $&42.1$\pm 0.4$ &24.7$\pm1.2$ &\textbf{47.5$\mathbf{\pm0.9}$ }  \\
        &F1-micro &25.8 $\pm 0.8$ & 27.3. $\pm0.7$ &43.2$\pm1.0$& 28.6 $\pm1.0$& 34.6 $\pm2.5$ &43.1$\pm0.3$ &25.4$\pm1.0$ &\textbf{48.4$\mathbf{\pm0.8}$ }  \\
        &AUC & 57.5 $\pm 0.5$& 58.2$\pm1.1$ &78.9$\pm 0.3$& 60.2$\pm1.0$& 65.6$\pm1.4$ &74.3$\pm0.0$ &55.0$\pm2.4$ &\textbf{79.4$\mathbf{\pm0.3}$ }  \\
    \midrule
        \multirow{3}*{Actor} 
        &F1-macro & 21.5$\pm 0.4$& 19.7 $\pm0.8$ &30.3$\pm4.7$ &28.0$\pm1.1$ &30.4$\pm1.1$ &\textbf{31.6}$\mathbf{\pm0.0}$ &20.0$\pm0.6$ &25.7$\pm0.44$   \\
        &F1-micro &29.2 $\pm0.6$ & 27.1 $\pm0.5$ &33.2$\pm0.6$ &32.5$\pm1.0$&33.7$\pm2.3$ &34.1$\pm0.0$ &25.5$\pm0.5$ &\textbf{34.7$\mathbf{\pm0.4}$ }  \\
        &AUC & 58.0 $\pm 0.5$& 55.8$\pm0.4$ &64.8$\pm0.1$ &56.2$\pm2.0$ &60.8.2$\pm0.6$ &64.7$\pm0.0$ &53.8$\pm0.1$ &\textbf{68.1$\mathbf{\pm0.2}$ }  \\
    \midrule
        ogbn-arxiv&Accuracy & 71.5$\pm 0.3$& 71.6$\pm0.5$ & 71.2 $\pm0.3$ & 71.7 $\pm0.1 $ & 71.9 $\pm0.6 $  &70.9$\pm0.2$ &70.0$\pm0.1$ & \textbf{72.2$\mathbf{\pm0.1}$}   \\
    \bottomrule
    \end{tabular}
    \caption{ Quantitative results (\%$\pm \sigma$) on node classification. (bold: best)}
    \label{table:baseline}
\end{table*}

\textbf{Baselines.}
The proposed HiD-Net is compared with several representative GNNs, including  three traditional GNNs: GCN \citep{gcn}, GAT \citep{gat}, APPNP \citep{appnp}, and four graph diffusion networks: GRAND \citep{grand}, GRAND++ \citep{grand++}, ADC \citep{adc}, DGC \citep{dgc}. They are implemented based on their open repositories, where the code can be found in Appendix.

\textbf{Experimental setting.}
We perform a hyperparameter search for HiD-Net on all datasets and the details of hyperparameter can be seen in Appendix.
For other baseline models: GCN, GAT, APPNP, GRAND, GRAND++, DGC, and ADC, we follow the
parameters suggested by \citep{gcn, gat, appnp, grand, grand++, dgc, adc} on Cora, Citeseer, and Pubmed, and carefully fine-tune them to get optimal performance on Chameleon, Squirrel, and Actor.
For all methods, we randomly run 5 times and report the mean and variance. More detailed experimental settings can be seen in Appendix.

\textbf{Results.} Table  ~\ref{table:baseline} summarizes the test results. Please note that OGB prepares standardized evaluators for testing results and it only provides accuracy metric for ogbn-arxiv. As can be seen, HiD-Net outperforms other baselines on seven datasets.
Moreover, in comparison with the graph diffusion networks, our HiD-Net is generally better than them with a large margin on heterophily graphs, which indicates that our designed graph diffusion process is more practical for different types of graphs.


\begin{figure*}[h]
\centering  
\subfigure{   
\begin{minipage}{\linewidth}
\centering    
\includegraphics[scale=0.3]{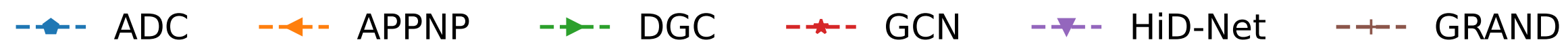}  
\end{minipage}
}

\subfigure[Cora]{   
\begin{minipage}{.32\linewidth}
\centering    
\includegraphics[scale=0.45]{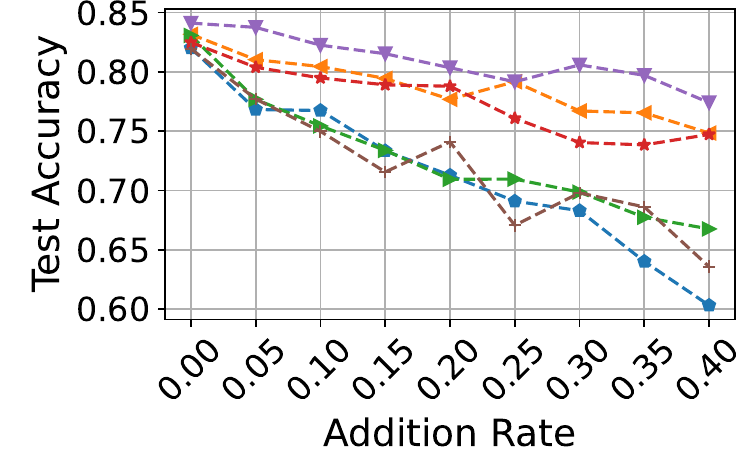}  
\end{minipage}
}
\subfigure[Citeseer]{ 
\begin{minipage}{.32\linewidth}
\centering    
\includegraphics[scale=0.45]{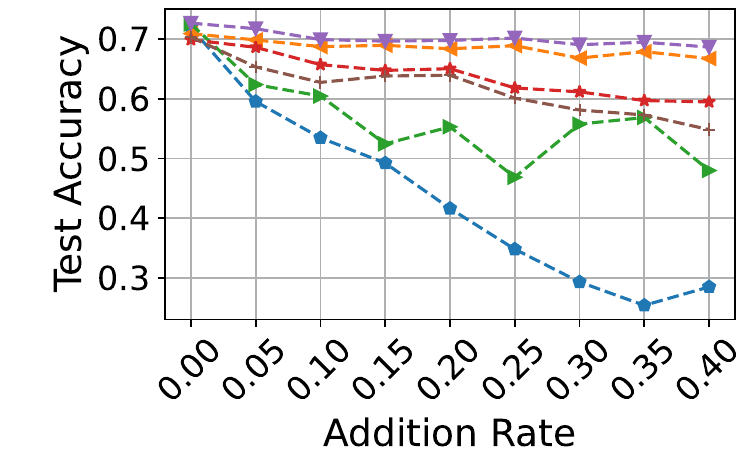}

\end{minipage}
}
\subfigure[Squirrel]{ 
\begin{minipage}{.32\linewidth}
\centering    
\includegraphics[scale=0.45]{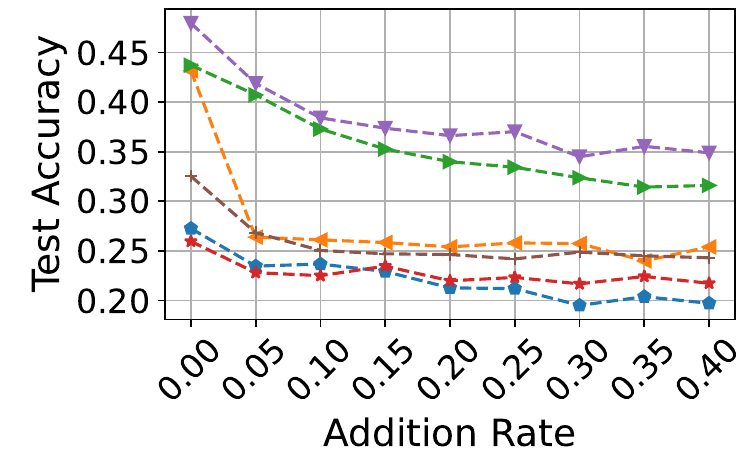}
\end{minipage}
}
\setlength{\abovecaptionskip}{0pt}
\caption{Results of different models under random edge addition.}    
\label{fig:attack_addition}    
\end{figure*}

\begin{figure*}[h]
\centering  
\subfigure[Cora]{   
\begin{minipage}{.32\linewidth}
\centering    
\includegraphics[scale=0.45]{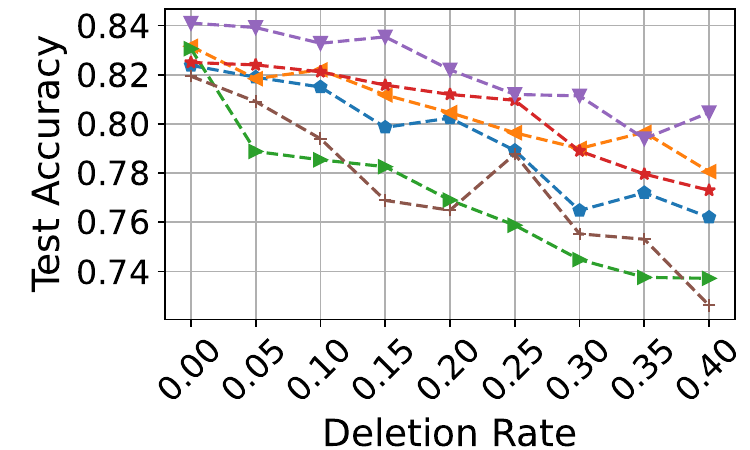}  

\end{minipage}
}
\subfigure[Citeseer]{ 
\begin{minipage}{.32\linewidth}
\centering    
\includegraphics[scale=0.45]{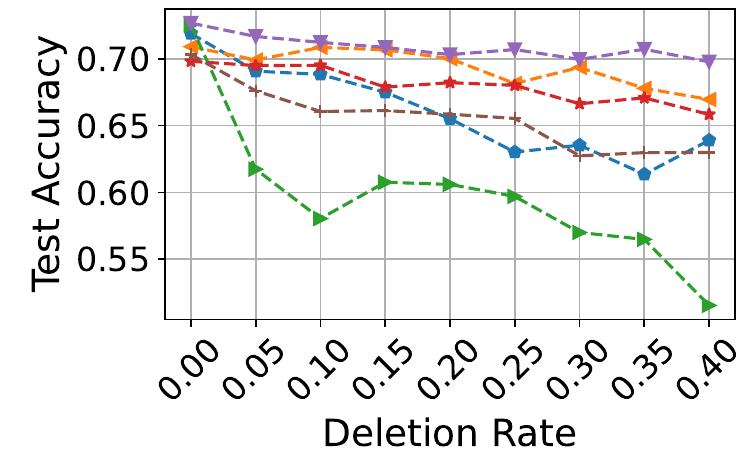}

\end{minipage}
}
\subfigure[Squirrel]{ 
\begin{minipage}{.32\linewidth}
\centering    
\includegraphics[scale=0.45]{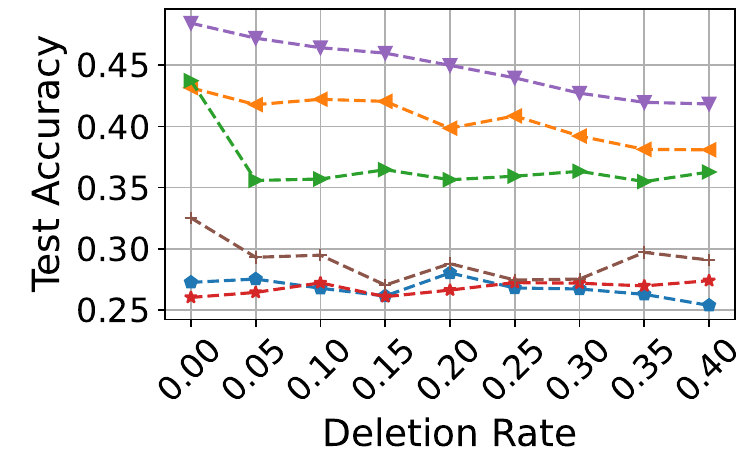}
\end{minipage}
}
\setlength{\abovecaptionskip}{0pt}
\caption{Results of different models under random edge deletion.}    
\label{fig:attack_deletion} 
\end{figure*}

\begin{figure*}[h]
\centering  
\subfigure[Cora]{   
\begin{minipage}{.32\linewidth}
\centering    
\includegraphics[scale=0.45]{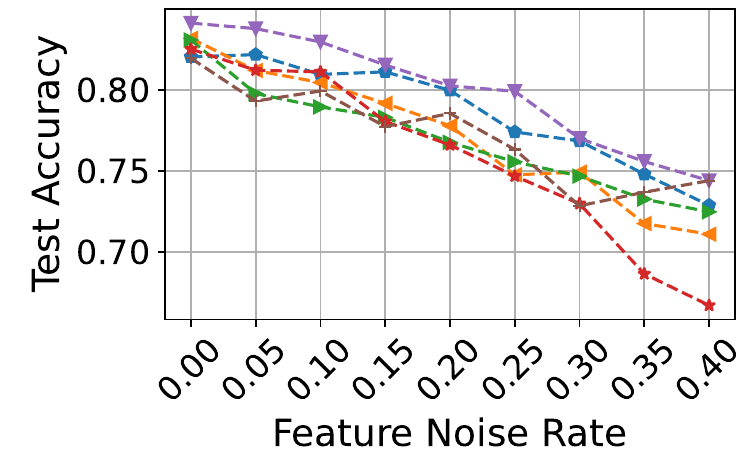}  

\end{minipage}
}
\subfigure[Citeseer]{ 
\begin{minipage}{.32\linewidth}
\centering    
\includegraphics[scale=0.45]{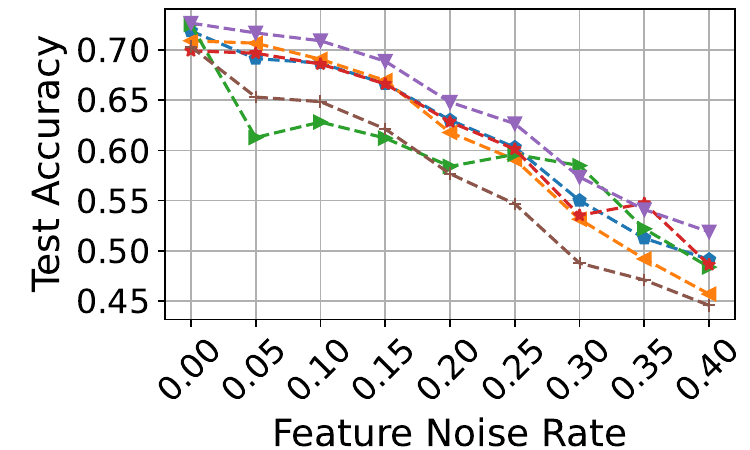}

\end{minipage}
}
\subfigure[Squirrel]{ 
\begin{minipage}{.32\linewidth}
\centering    
\includegraphics[scale=0.45]{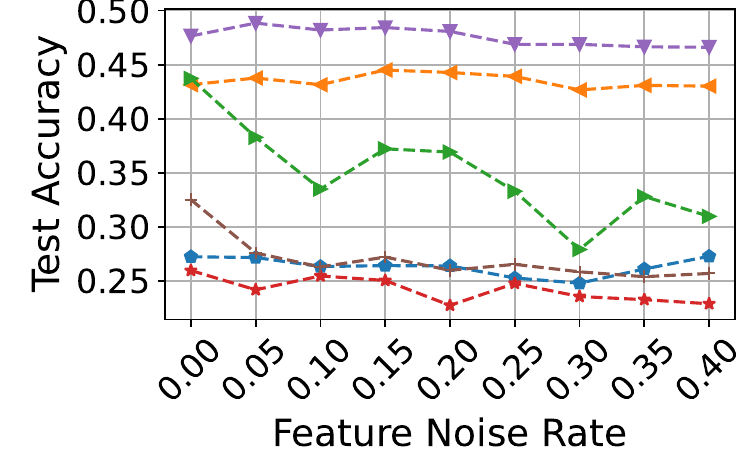}
\end{minipage}
}
\setlength{\abovecaptionskip}{0pt}
\caption{Results of different models under random feature perturbation.}
    \label{fig:attack_feature}
\end{figure*}

\begin{figure}[h]
\centering  

\subfigure[Cora]{   
\begin{minipage}{.45\linewidth}
\flushleft 
\includegraphics[scale=0.32]{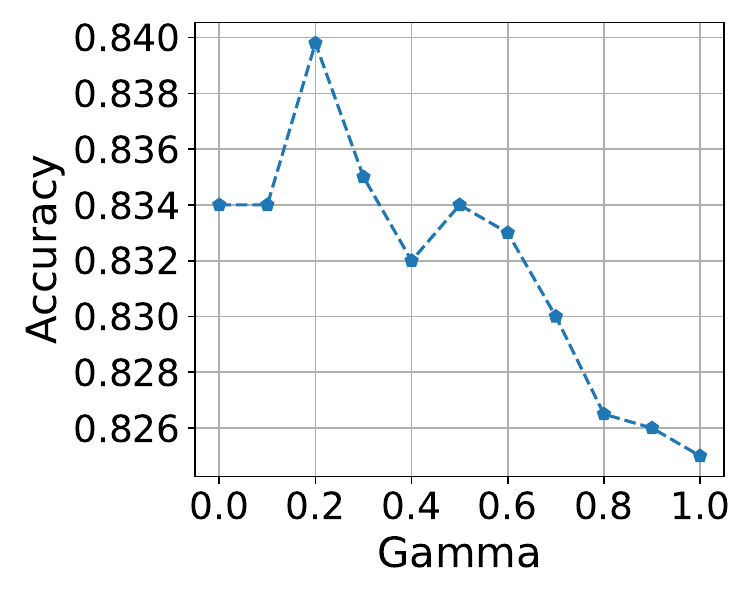}  
\label{fig:rcora}
\end{minipage}
}
\subfigure[Chameleon]{ 
\begin{minipage}{.45\linewidth}
\centering    
\includegraphics[scale=0.32]{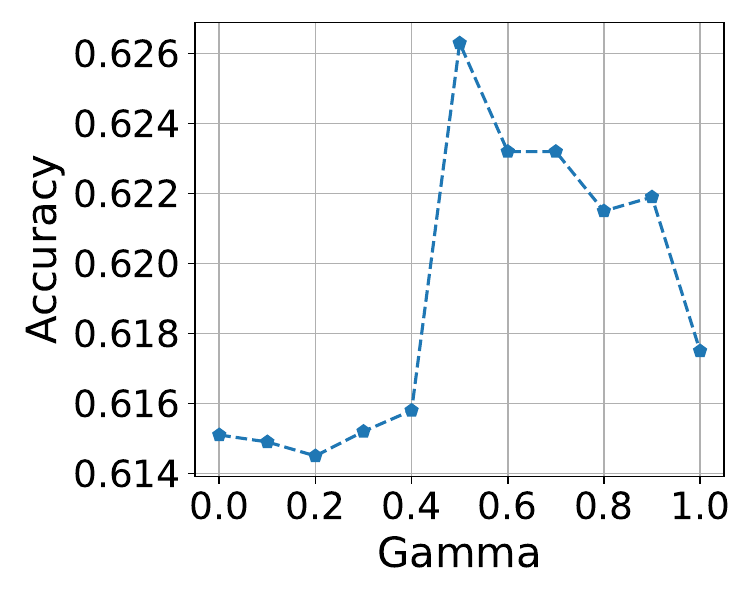}
\label{fig:rciteseer}
\end{minipage}
}
\setlength{\abovecaptionskip}{0pt}
\caption{Analysis of parameter $\gamma$.}    

\label{fig:r}    
\end{figure}

\begin{figure}[h]
\centering
\vspace{-2mm}
\hspace{-10mm}
\setlength{\abovecaptionskip}{0.cm}
\subfigure[Chameleon]{
\begin{minipage}[t]{0.42\linewidth}
\includegraphics[width=1.2\columnwidth]{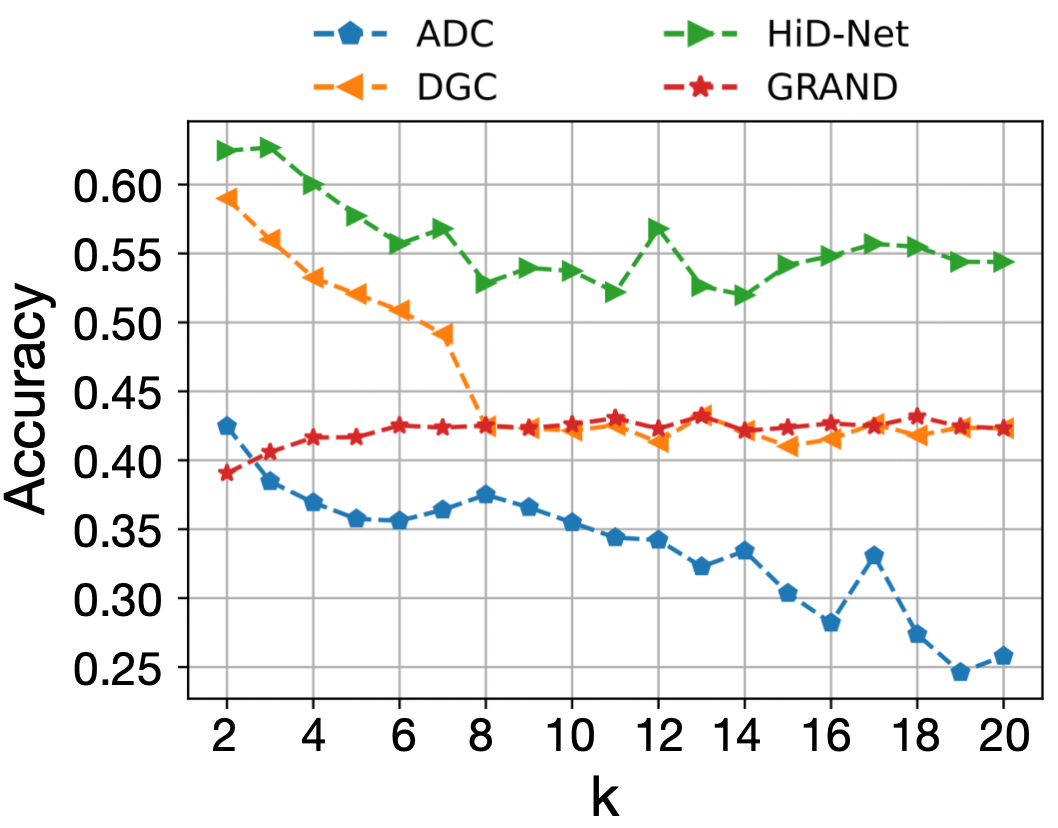}
\end{minipage}}\hspace{5mm}
\subfigure[Actor]{
\begin{minipage}[t]{0.42\linewidth}
\includegraphics[width=1.2\columnwidth]{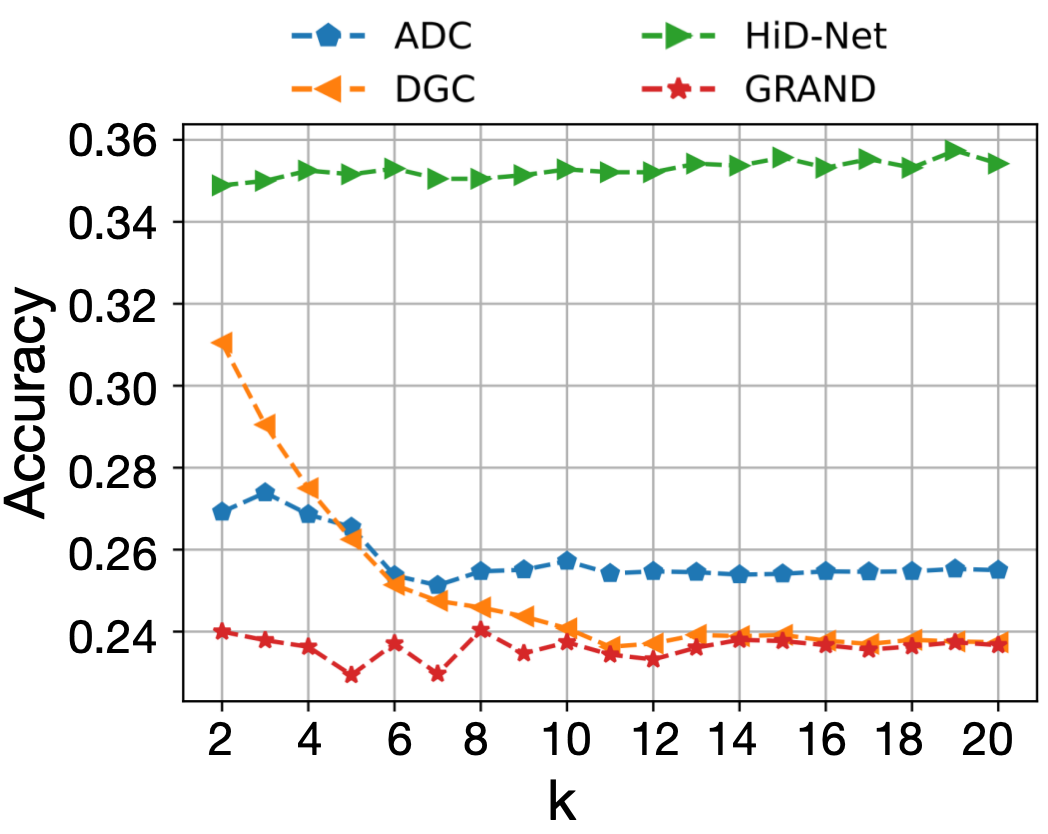}
\end{minipage}
}
\caption{Non-over-smoothing with increasing steps.} \label{fig:over} 
\end{figure}

\begin{figure}[h]
\centering
\hspace{-12mm}
\setlength{\abovecaptionskip}{0.cm}
\subfigure[$\alpha$]{
\setlength{\abovecaptionskip}{0.cm}
\begin{minipage}[t]{0.42\linewidth}
\includegraphics[width=1.2\columnwidth]{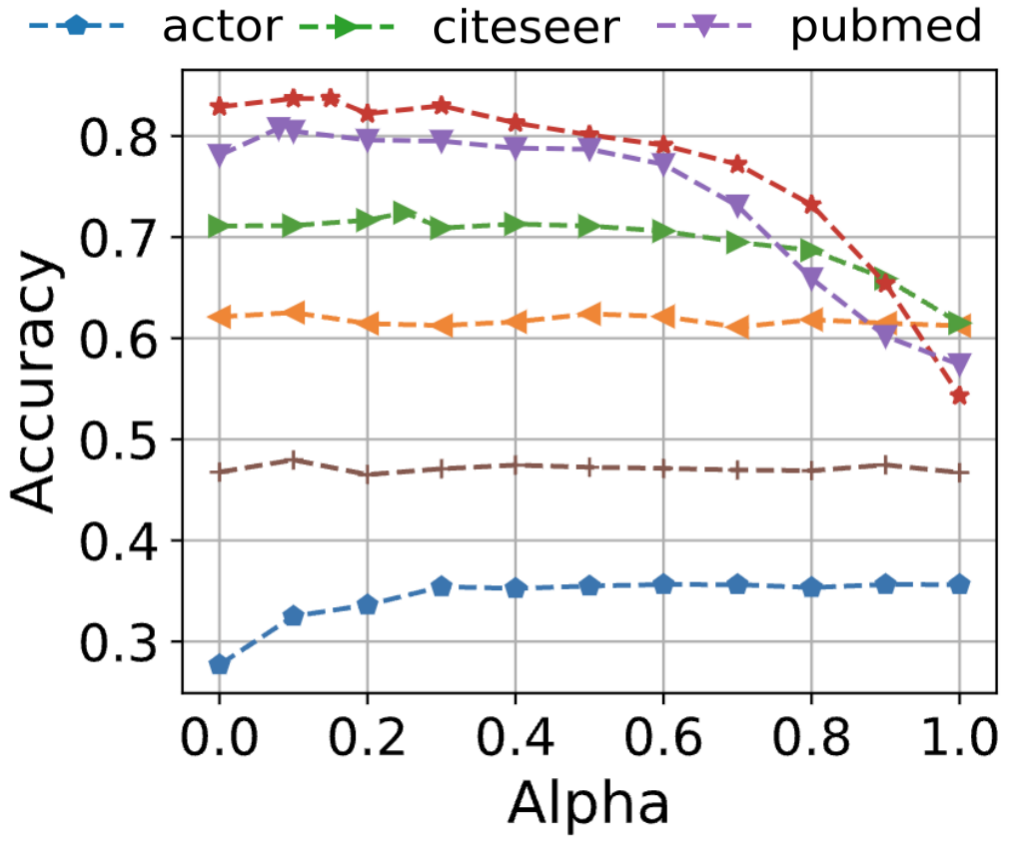}
\label{fig:aba}
\end{minipage}}\hspace{5mm}
\subfigure[$\beta$]{
\setlength{\abovecaptionskip}{0.cm}
\begin{minipage}[t]{0.42\linewidth}
\includegraphics[width=1.2\columnwidth]{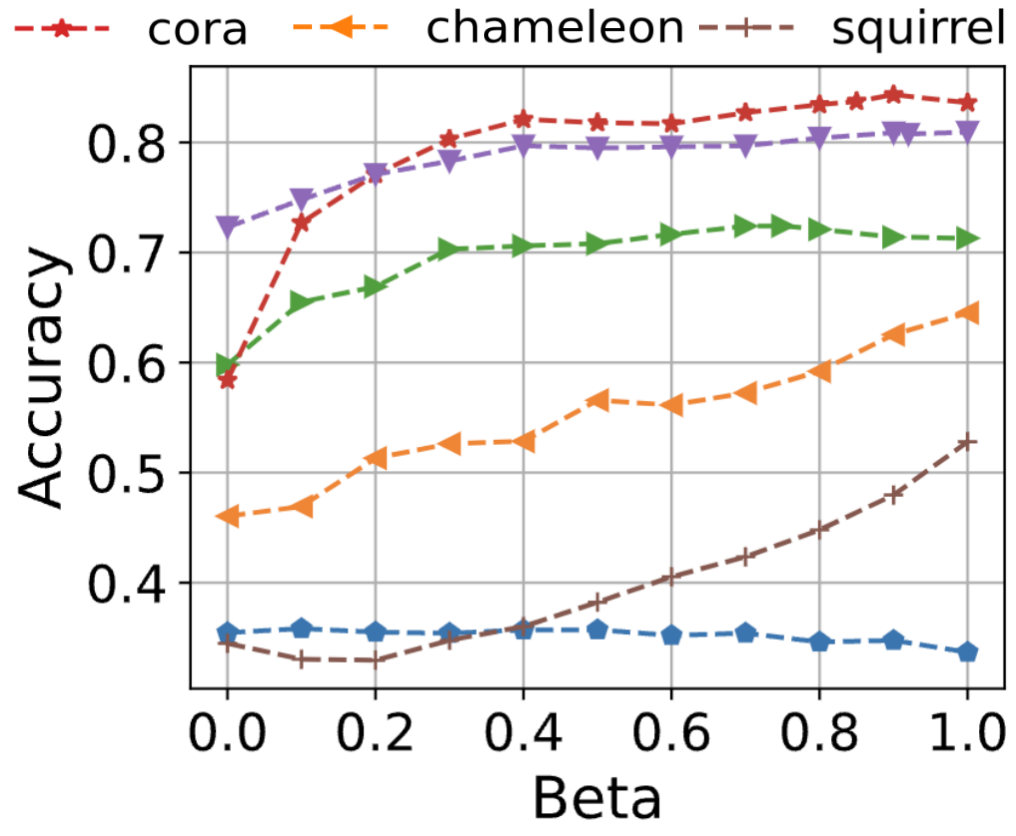}
\label{fig:abb}
\end{minipage}}
\caption{Analysis of parameter $\alpha$ and $\beta$.}\label{fig:ab} 
\end{figure}


\subsection{Robustness Analysis}
Utilizing the information from 2-hop neighbors, our model is more robust in abnormal situations. We comprehensively evaluate the robustness of our model on three datasets (Cora, Citeseer and Squirrel) in terms of attacks on edges and features, respectively.

\vspace{-0mm}
\textbf{Attacks on edges.}
To attack edges, we adopt random edge deletions or additions following \citep{attack1, attack2}. For edge deletions and additions, we randomly remove or add 5\%, 10\%, 15\%, 20\%, 25\%, 30\%, 35\%, 40\% of the original edges, which retains the connectivity of the attacked graph. Then we perform node classification task. All the experiments
are conducted 5 times and we report the average accuracy. The
results are plotted in Figure ~\ref{fig:attack_addition} and Figure ~\ref{fig:attack_deletion}. From the figures, we can see that as the addition or deletion rate rises, the performances on three datasets of all the models degenerate, and HiD-Net consistently outperforms other baselines.

\vspace{-0.4mm}
\textbf{Attacks on features.}
To attack features, we inject random perturbation into the node features as in \citep{attack3}. Firstly, we sample a noise matrix $\mathbf{M} \in \mathbb{R} ^{n \times q}$, where each entry in $\mathbf{M}$ is sampled from the normal distribution $N(0, 1)$. Then, we calculate reference amplitude $r$, which is the mean of the maximal value of each node’s feature. 
We add Gaussian noise $\mu \cdot r \cdot \mathbf{M}$ to the original feature matrix, and get the attacked feature matrix, where $\mu$ is the noise ratio. The results are reported in Figure ~\ref{fig:attack_feature}. Again, HiD-Net consistently outperforms all other baselines under different perturbation rates by a margin for three datasets.

\subsection{Non-over-smoothing with Increasing Steps}

To demonstrate that our model solves the oversmoothing
problem compared with other graph diffusion networks, we test different graph diffusion models with increasing propagation step $k$ from 2 to 20. Baselines include DGC, ADC and GRAND. The results are plotted in Figure ~\ref{fig:over}. We can see that with the increase of $k$, HiD-Net consistently performs better than other baselines.

\subsection{Parameter Study}

In this section, we investigate the sensitivity of parameters on all datasets.

\textbf{Analysis of $\alpha$.} We test the effect of $\alpha$ in  ~\eqref{eq:model},  and vary it from 0 to 1. From Figure ~\ref{fig:aba} we can see that with the increase of $\alpha$, the performances of citation graphs rise first and then start to drop slowly, the performances of Chameleon and Squirrel have not changed too much, and the performance of Actor first rises and then remains unchanged. As citation graphs are more homophily, we need to focus less on the node itself, implying a small $\alpha$, while on heterophily graphs, we need to focus more on the node itself.

\textbf{Analysis of $\beta$.} In order to check the impact of the diffusion term, we study the performance of HiD-Net with $\beta$ varying from 0 to 1. The results are shown in Figure ~\ref{fig:abb}. We can see that as the value of $\beta$ increases, the accuracies generally increase, while the accuracy on Actor remains relatively stable, implying that the features diffused from 1-hop and 2-hop neighbors are very informative.

\textbf{Analysis of $\gamma$.} Finally we test the effect  of $\gamma$ in ~\eqref{eq:model} and vary it from 0  to  0.6. With the increase of $\gamma$, the accuracies on different datasets do not change much, so we just separately plot each dataset for a clearer illustration here. 
As can be seen in Figure ~\ref{fig:r}, with the increase of $\gamma$, the performance on Cora and Chameleon rises first and then drops, and different graphs have different best choices of $\gamma$.
The results on other datasets are shown in Appendix.

\vspace{-2mm}

\section{Conclusion}
\label{section:conclusion}
In this paper, we propose a generalized diffusion graph framework, which establishes the relation between diffusion equation with different GNNs. Our framework reveals that current 
graph diffusion networks mainly consider the first-order diffusion equation, then based on our finding of the monophily property of labels, we derive a novel high-order diffusion graph network (HiD-Net). HiD-Net is more robust and general on both homophily and heterophily graphs.  Extensive experimental results verify the effectiveness of HiD-Net. One potential issue is that our model utilizes a constant diffusivity coefficient, and a future direction is to explore a learnable diffusivity coefficient depending on time and space. Our work formally points out the relation between diffusion equation with a wide variety of GNNs. Considering that previous GNNs are designed mainly based on spatial or spectral strategies, this new framework may open a new path to understanding and deriving novel GNNs. We believe that more insights from the research community on the diffusion process will hold great potential for the GNN community in the future. 

\vspace{-2mm}
\section{Acknowledgments}
This work is supported in part by the National Natural Science Foundation of China (No. U20B2045, 62192784, U22B2038, 62002029, 62172052, 62322203).

\newpage

\bibliography{aaai24}

\newpage

\newpage
\clearpage

\appendix

 \begin{table*}[h]
    \centering
    \setlength{\belowcaptionskip}{0mm}
    
    \begin{tabular}{c|c|c|c|c|c|c|c|c|c}
    \toprule
        Dataset & $\alpha$ & $\beta$ & $\gamma$  & $\Delta t$ & k &Hidden dimension &Learning rate & Weight Decay & dropout\\
    \midrule
    Cora&0.1 &0.9 &0.3&0.8 &10 &128 &0.01 &0.00 & 0.55\\
    Citeseer&0.1 &0.9 & 0.2 &0.6& 10&64 & 0.005& 0.05& 0.5\\
    Pubmed&0.08 &0.92 &0.3 &1&8 &32 & 0.02&0.0005&0.5  \\
    Chameleon&0.1 &0.9 &0.05 &1&1 &64 & 0.6&0.01 & 0.0005\\
    Squirrel& 0.1&0.9 &0.3 &1&1 &64 &0.02 &0.0005 &0.5\\
    Actor&0.1 &0.9 &0.1&0.2 &2 &64 &0.005 &0.005&0.5 \\
    ogbn-arxiv&0.05 &0.94 &0.1 &0.9 &4 &64 &0.02 &0.0005&0.5 \\
        \bottomrule
    \end{tabular}
    \caption{ Hyperparameters for HiD-Net}
    \label{table:parameter}
\end{table*}

\section{Proofs}
\label{appendix:proofs}

\subsection{Proof of Proposition 1}

Graph Convolutional Network (GCN) has the following propagation mechanism which conducts linear transformation and nonlinearity activation repeatedly throughout $K$ layers:
\begin{equation}
    \mathbf{X}^{(K)}=\sigma\left(\hat{\tilde{\mathbf{A}}}\left(\cdots\left(\sigma\left(\hat{\tilde{\mathbf{A}}} \mathbf{X}^{(0)} \mathbf{W}^{(0)}\right) \cdots\right) \mathbf{W}^{(K-1)}\right)\right),
\end{equation}

where $\mathbf{W}^{(i)}$ is the trainable weight matrix at layer $i$. Simplifying Graph Convolutional Network (SGC) \citep{sgc} removes nonlinearities and collapses weight matrices between consecutive layers, which has similar propagation mechanism with GCN and exhibits comparable performanceas. SGC propagates as:

\begin{equation}
\mathbf{X}^{(K)}=\hat{\tilde{\mathbf{A}}} \cdots \hat{\tilde{\mathbf{A}}} \mathbf{X}^{(0)} \mathbf{W}^{(0)} \mathbf{W}^{(1)} \ldots \mathbf{W}^{(K-1)}=\hat{\tilde{\mathbf{A}}}^{(K)} \mathbf{X} \mathbf{W}^{*},
\end{equation}
where $\mathbf{W}^{*}=\mathbf{W}^{(0)} \mathbf{W}^{(1)} \cdots \mathbf{W}^{(K-1)}$.  We show that the propagation mode of SGC (GCN) is a special form of the generalized diffusion framework:

According to ~\eqref{eq:sgc} we have:
\begin{equation}
\dfrac{\partial x_i^{(t)}}{\partial t}= \operatorname{div} ((\nabla x)_{ij}^{(t)})=\sum_{j\in N_1(i)}\dfrac{\tilde{A_{ij}}}{\sqrt{\tilde{d_i}} \sqrt{\tilde{d_j}}}(x_j - x_i).
\end{equation}

With $\Delta t = 1$, on all nodes we have:

\begin{equation}
\mathbf{X}^{(t+1)} - \mathbf{X}^{(t)} = (\mathbf{\hat{\tilde{A}}}-\mathbf{I})\mathbf{X}^{(t)}= -\mathbf{\tilde{L}} \mathbf{X}^{(t)},
\end{equation}
so
\begin{equation}
\mathbf{X}^{t+1}= \mathbf{\hat{\tilde{A}}}\mathbf{X},
\end{equation}
which matches the propagation mechanism of SGC.

\subsection{Proof of Proposition 2.}

PPNP is a graph neural network which utilizes a propagation scheme derived from personalized PageRank and adds a chance of teleporting back to the root node, which ensures that the PageRank score encodes the local neighborhood for every root node. PPNP’s model equation is:

\begin{equation}
\mathbf{X}=\operatorname{softmax}\left(\eta\left(\boldsymbol{I}-(1-\eta) \hat{\tilde{\mathbf{A}}}\right)^{-1} \mathbf{X}^{(0)}\right),
\end{equation}
and APPNP approximates PPNP via power iteration:

\begin{equation}
\mathbf{X}^{(k+1)} =(1-\eta) \hat{\tilde{\mathbf{A}}} \mathbf{X}^{(k)}+\eta \mathbf{X}^{(0)}.
\end{equation}

As stated in \textbf{Remark 1}, the steady state of ~\eqref{eq:diffusion} is ~\eqref{eq:euler}. With $\alpha=1$, $\beta=1-\dfrac{1}{\eta}$, $\Delta t = 1$  and $f=1$ in  ~\eqref{eq:euler}, on all nodes, we have:
\begin{equation}
0 = (\mathbf{X}-\mathbf{X}^{(0)})+(1-\dfrac{1}{\eta})(\mathbf{\hat{\tilde{A}}}-\mathbf{I})\mathbf{X}.
\end{equation}

So
\begin{equation}
\begin{aligned}
    \mathbf{X}
    &=(\mathbf{I} +(1-\dfrac{1}{\eta}) (\hat{\tilde{\mathbf{A}}}-\mathbf{I}))^{-1}  \mathbf{X}^{(0)}\\
    &=\eta (\mathbf{I} - (1-\eta) \hat{\tilde{\mathbf{A}}})^{-1} \mathbf{X}^{(0)},
\end{aligned}
\end{equation}
which works the same as PPNP.

\subsection{Proof of Proposition 3.}

GAT \citep{gat} is an attention-based architecture to perform node classification of graph-structured data. The attention coefficient between node $i$ and node $j$ is computed based on a shared attentional mechanism:

\begin{equation}
f_{i j}^{(t)}=\frac{\exp \left(\operatorname{LeakyReLU}\left({\mathbf{a}}^{(t)}\left[\mathbf{W} x_i^{(t)} \| \mathbf{W} x_j^{(t)}\right]\right)\right)}{\sum_{k \in \mathcal{N}_{i}} \exp \left(\operatorname{LeakyReLU}\left({\mathbf{a}}^{(t)}\left[\mathbf{W} x_i^{(t)} \| \mathbf{W} x_k^{(t)}\right]\right)\right)},
\end{equation}
where $\mathbf{a}^{(t)}$ is a shared attentional mechanism at time $t$, and $\mathbf{W}$ is the weight matrix.

As in \eqref{eq:gat}, with $\Delta t = 1$, we have:
\begin{equation}
    \dfrac{\partial x_i^{(t)}}{\partial t}=\sum_{j\in N_1(i)} f_{ij}^{(t)}(x_j^{(t)}-x_i^{(t)}).
\end{equation}
    
Let $\mathbf{F}^{(t)}$ be the attention matrix at time $t$, and $f_{ij}^{(t)}$ is the element of it $\mathbf{F}^{(t)}$. With $\Delta t = 1$, on all nodes, we have
    \begin{equation}
        \mathbf{X}^{(t+1)} - \mathbf{X}^{(t)}=
     (\mathbf{F}^{(t)} - \mathbf{I}) \mathbf{X}^{(t)},
    \end{equation}
    so,
    \begin{equation}
        \mathbf{X}^{(t+1)} =
     \mathbf{F}^{(t)} \mathbf{X}^{(t)},
\end{equation}
which works as the propagation of GAT.

\subsection{Proof of Proposition 4.}

AMP \citep{amp} designs a better message passing scheme with node-wise adaptive feature aggregation and residual connection. With the normalized ajacency matrix, it propagates as:

\begin{equation}
\left\{\begin{aligned}
\mathbf{Y}^{(t)} &=(1-2 \epsilon(1-\lambda)) \mathbf{X}^{(t)}+2 \epsilon(1-\lambda) \hat{\tilde{\mathbf{A}}} \mathbf{X}^{(t)} \\
\beta_{i}^{(t)} &=\max \left(1-\frac{\epsilon \lambda}{\left\|\mathbf{Y}_{i}^{(t)}-x^{(0)}_{i}\right\|_{2}}, 0\right) \quad \forall i \in[n], \\
x_{i}^{(t+1)} &=\left(1-\beta_{i}\right)x^{(0)}_{i}+\beta_{i} \mathbf{Y}_{i}^{(t)} \quad \forall i \in[n]
\end{aligned}\right.
\end{equation}
where $\epsilon$ is the stepsize, and $\lambda$ is the coefficient.

From ~\eqref{eq:amp}, with $\Delta t = 1$ we have:

\begin{equation}
\begin{aligned}
    &x_i^{(t+1)}-x_i^{(t)} = (1-\beta_i^{(t)}) (x_i^{(0)} - x_i^{(t)}) \\&+ 2\epsilon(1-\lambda) \beta_i^{(t)} \operatorname{div}( (\nabla x)_{ij}^{(t)}),
\end{aligned}
\end{equation}
so
\begin{equation}
\begin{aligned}
 x_i^{(t+1)}&= (1-\beta_i^{(t)}) x_i^{(0)} + (1-2\epsilon(1-\lambda)) \beta_i^{(t)}x_i^{(t)}\\&+ 2\epsilon(1-\lambda) \beta_i^{(t)}(\hat{\tilde{\mathbf{A}}}\mathbf{X})_i^{(t)}\\&=\left(1-\beta_{i}\right)x^{(0)}_{i}+\beta_{i} \mathbf{Y}_{i}^{(t)},
\end{aligned}
\end{equation}
which works as the propagation of AMP.

\subsection{Proof of Proposition 5.}

Deep Adaptive Graph Neural Networks (DAGNN) \citep{dagnn} adaptively incorporates information from large receptive fields. DAGNN propagates as:

\begin{equation}
\begin{aligned}
 \mathbf{X^K} 
&=s_{0} \mathbf{X}^{(0)}+s_{1} \hat{\tilde{\mathbf{A}}} \mathbf{X}^{(0)}+s_{2} \hat{\tilde{A}}^{2} \mathbf{X}^{(0)}+\cdots+s_{K} \hat{\tilde{\mathbf{A}}}^{K} \mathbf{X}^{(0)} \\
&=\sum_{k=0}^{K} s_{k} \hat{\tilde{\mathbf{A}}}^{k} \mathbf{X}^{(0)},    
\end{aligned}
\end{equation}

where $s_{0}, s_{1}, \cdots, s_{K}$  are the learnable retainment coefficients and $\sum_{k=0}^{K} s_{k}=1$.

When $t\rightarrow \infty$, we have:

\begin{equation}
\begin{aligned}
    \mathbf{X}^{(t)}&=\alpha(\mathbf{I}-\beta f(t)\hat{\tilde{\mathbf{A}}})\mathbf{X}^{(0)}\\
    &=\alpha \sum_{t=0}^{T}(\beta f(t)\hat{\tilde{\mathbf{A}}})^{t}\mathbf{X}^{(0)},
\end{aligned}
\end{equation}
as the propagation of DAGNN goes.

\subsection{Proof of Proposition 6}

\begin{proof}

Clearly, for all for all $i = 1, \cdots, n$

\begin{equation}
    \mathbb{E}(x_i^{(0)}) = x_i^{(0)}.
\end{equation}

for all for all $i = 1, \cdots, n$, assume that

\begin{equation}
    \mathbb E(x_i^{(t)}) = x_i^{(t)}.
\end{equation}

Then, for all $i = 1, \cdots, n$, we have

\begin{equation}
\begin{aligned}
    \mathbb{E}(\mathbb{X}_i^{t+\Delta t}) &= \sum_{j \in V} P_{ji} \mathbb{E}(x_j^{(t)})\\
    &=\alpha \Delta t \mathbb{E}(x_i^{(0)}) + \left[1-(\alpha + \beta) \Delta t\right]\mathbb{E}(x_i^{(t)}) \\&+ (\beta - \gamma)\Delta t \sum_{j\in N_1(i)} \hat{\tilde{A}}_{ij}\mathbb{E}(x_j^{(t)})\\&+ \gamma \Delta t \sum_{j\in N_2(i)} \hat{\tilde{A}}^2_{ij} \mathbb{E}(x_j^{(t)})\\
    &=\alpha \Delta t x_i^{(0)} + \left[1-(\alpha + \beta) \Delta t\right]x_i^{(t)} \\&+ (\beta - \gamma)\Delta t \sum_{j\in N_1(i)} \hat{\tilde{A}}_{ij}x_j^{(t)}\\&+ \gamma \Delta t \sum_{j\in N_2(i)} \hat{\tilde{A}}^2_{ij} x_j^{(t)}\\
    &=x_i^{t+\Delta t}.
\end{aligned}
\end{equation}

as required.

\end{proof}

\subsection{Proof of Proposition 7.}

From ~\eqref{eq:model}, on all nodes, we have
    
\begin{equation}
\begin{aligned}
    &\mathbf{X}^{t+\Delta t}=\alpha \Delta t \mathbf{X}^{(0)}  + \left[ 1-(\alpha + \beta)\Delta t \right. \\
    & \left. + (\beta - \beta \gamma) \Delta t \hat{\tilde{\mathbf{A}}} +\beta \gamma \Delta t\hat{\tilde{\mathbf{A}}}^2 \right] \mathbf{X}^{(t)}.
\end{aligned}
\label{eq:diffusion7}
\end{equation}
    
    Then we have
    
\begin{equation}
    \setlength{\abovedisplayskip}{3pt}
\setlength{\belowdisplayskip}{3pt}
        \mathbf{X}^{t+\Delta t}=\alpha \Delta t \mathbf{X}^{(0)} + \mathbf{C} \mathbf{X}^{(t)},
\end{equation}
where $\mathbf{C} = \left[ \mathbf{I}-(\alpha + \beta)\Delta t \mathbf{I} + (\beta - \beta \gamma) \Delta t \hat{\tilde{\mathbf{A}}} + \beta \gamma \Delta t\hat{\tilde{\mathbf{A}}}^2 \right]$.
    
    The resulting prediction at time $t$ is
    
    \begin{equation}
        \mathbf{X}^{(t)} = (\mathbf{C}^{t} + \alpha \Delta t \sum_{i=0}^{t-\Delta t}\mathbf{C}^i)\mathbf{X}^{(0)}.
        \label{eq:limit}
    \end{equation}
    
    Since $f=1$, $\alpha, \beta, \gamma, \Delta t \in \left(0, 1\right]$ and $\hat{\tilde{\mathbf{A}}}$
is symmetrically normalized, we have $\operatorname{det}(\hat{\tilde{\mathbf{A}}}), \operatorname{det}(\hat{\tilde{\mathbf{C}}}) \le 1$, taking the limit $t \rightarrow \infty$ the left term tends to 0 and the right term becomes a geometric series. Resulting in

\begin{equation}
\begin{aligned}
\mathbf{X}^{\infty} &= \alpha\Delta t (\mathbf{I} - \mathbf{C})^{-1} \mathbf{X}^{(0)}\\
&=\alpha((\alpha + \beta) \mathbf{I} - \beta(1- \gamma)  \hat{\tilde{\mathbf{A}}} -\beta \gamma \hat{\tilde{\mathbf{A}}}^2)^{-1} \mathbf{X}^{(0)}.
\end{aligned}
\end{equation}

\subsection{Proof of Proposition 8.}

Firstly we assume that there exists $\bar{x}$ and $x_i^{(t)} \rightarrow \bar{x}$ for all $i \in V$ when $t\rightarrow \infty$, then $ \operatorname{div} (f(\nabla x_i^{(t)} ))=0$ and $ \operatorname{div} (\overline{(\nabla x)_i^{(t)}})=0$, so
    
    \begin{equation}
    \begin{aligned}
    &x_{i}^{(t+\Delta t)}-x_{j}^{(t+\Delta t)}=\alpha \Delta t\left(x_{i}^{(0)}-x_{j}^{(0)}\right)\\&+(1-\alpha \Delta t)\left(x_{i}^{(t)}-x_{j}^{(t)}\right)   \rightarrow \alpha\left(x_{i}^{(0)}-x_{j}^{(0)}\right) \neq 0,
    \end{aligned}
    \end{equation}
    which contradict to the assumption. As $t \rightarrow \infty$, the features among neighbors are less likely to be similar, thus alleviating the oversmoothing issue.

\section{Reproducibility Information}

\begin{table*}[t]
    \centering
    \begin{tabular}{c|ccccc}
    \toprule
        Dataset & Nodes & Edges & Classes & Features & Train/Val/Test nodes\\
    \midrule
    Cora& 2,708 & 5,429 & 7 & 1,433 & 140/500/1,000\\
    Citeseer& 3,327 & 4,732 & 6 & 3,703 &120/500/1,000\\
    Pubmed&19,717&44,338 & 3 & 500 & 60/500/1,000\\
    ogbn-arxiv& 169,343&1,166,243& 40& 128& 90941/29799/48603\\
    \midrule
    Chameleon&2,277 &36,101 &5 &2,325 &1092/729/456\\
    Squirrel&5,201&217,073 &5 &2,089 &2496/1664/1041\\
    Actor&7,600 &33,544 &5 &931 &3648/2432/1520 \\
    
        \bottomrule
    \end{tabular}
    \caption{ Description of datasets.}
    \label{table:datasets}
\end{table*}

\label{appendix:experiment}

\subsection{Datasets}

The details of these datasets are summarized in Table ~\ref{table:datasets}.

\subsection{Hyperparameters}

We use a two-layer MLP as the model $l_{\omega}(\cdot)$, following APPNP. For all models, we use the Adam optimizer and search the optimal learning rate over \{0.005, 0.01, 0.02, 0.1, 0.6\}, weight decay \{0.0, 0.0005, 0.005, 0.01, 0.05\}, hidden dimension \{32, 64, 128\}, dropout \{0.0005, 0.1, 0.5, 0.55\}, $k$ \{1, 2, 4, 8, 10\}, $\Delta t$ \{0.2, 0.4, 0.6, 0.8, 0.9, 1\}, $\alpha$ \{0.05, 0.08, 0.1\}, $\beta$ \{0.9, 0.92, 0.94\}, $\gamma \{0.05, 0.1, 0.2, 0.3\}$.
 
The hyperparameters are shown in Table ~\ref{table:parameter}.

\subsection{Environment}
 The environment where our code runs is shown as follows:
\begin{itemize}
    \item Operating system: Linux version 3.10.0-693.el7.x86\_64
    \item CPU information: Intel(R) Xeon(R) Silver 4210 CPU @ 2.20GHz
    \item GPU information: GeForce RTX 3090
\end{itemize}

    
\subsection{Other Source Code}

The acquisition of all the code below complies with the provider’s license and do not contain
personally identifiable information and offensive content. The address of code of baselines are listed
as follows:

GCN (MIT license): \url{https://github.com/tkipf/pygcn}

GAT (MIT license): \url{https://github.com/Diego999/pyGAT}

APPNP (MIT license): \url{https://github.com/klicperajo/ppnp}

GRAND (MIT license): \url{https://github.com/twitter-research/graph-neural-pde}

ADC (MIT license): \url{https://github.com/abcbdf/ADC}

DGC (MIT license): \url{https://github.com/yifeiwang77/dgc}


        
         
        

            


            
\section{Time complexity}
\label{appendix:complexity}
The propagation step of HiD-Net can be formulated as:HiD-Net can be formulated as:

\begin{align*}
 &\mathbf{X}^{(t+\Delta t)}=\alpha \Delta t \mathbf{X}^{(0)} + \left[ 1-(\alpha + \beta)\Delta t + (\beta - \beta \gamma) \Delta t \hat{\tilde{\mathbf{A}}} \right. \\ &\left.+\beta \gamma \Delta t\hat{\tilde{\mathbf{A}}}^2 \right] \mathbf{X}^{(t)}   
\end{align*}

As $\hat{\tilde{\mathbf{A}}}^2$ can be precomputed, it takes $\mathcal{O}(n^2\zeta)$ to calculate $\hat{\tilde{\mathbf{A}}}^2\mathbf{X}$, where $\mathbf{X}\in\mathbb{R}^{n \times \zeta}$, $n$ is the number of the nodes, $\zeta$ is the dimension of the feature vector. It takes $\mathcal{O}(n^2\zeta)$ to calculate $\hat{\tilde{\mathbf{A}}}\mathbf{X}$, and $\mathcal{O}(1)$ to calculate $\alpha \Delta t \mathbf{X}^{(0)}$ and $(1-(\alpha + \beta)\Delta t)\mathbf{X}^{(t)}$. So the time complexity of the propagation step of HiD-Net is  $\mathcal{O}(n^2\zeta)$ and the first-order diffusion is also $\mathcal{O}(n^2\zeta)$. 

\section{More results}
\label{Appendix:figures}
\begin{figure*}[h]
\centering  

\subfigure[Squirrel]{   
\begin{minipage}{.23\linewidth}
\centering    
\includegraphics[scale=0.25]{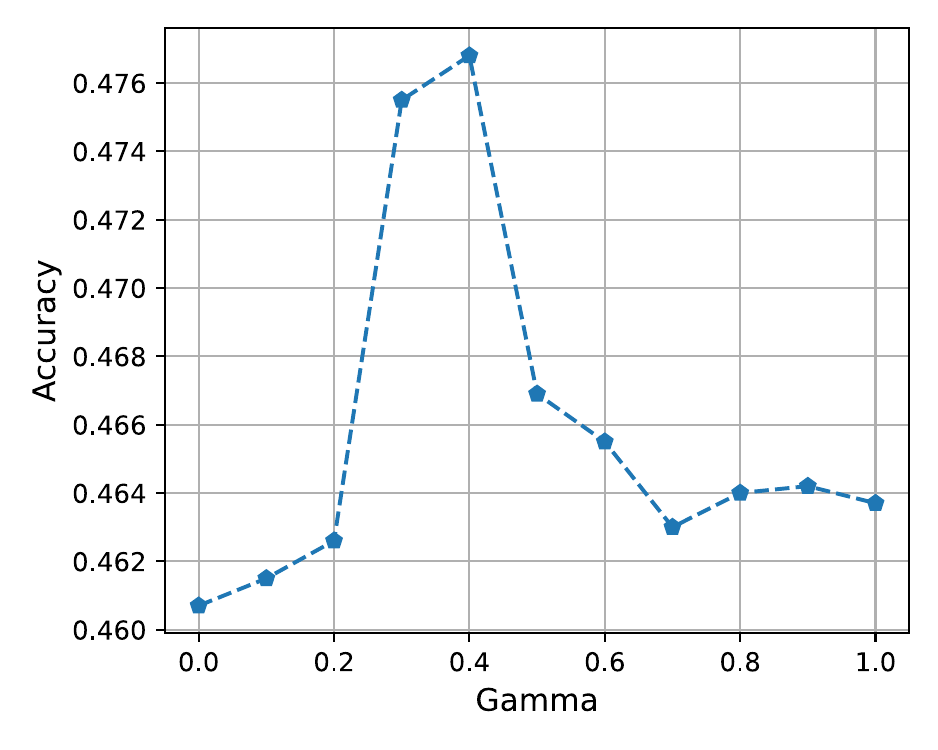}  
\label{fig:rcora}
\end{minipage}
}
\subfigure[Pubmed]{   
\begin{minipage}{.23\linewidth}
\centering    
\includegraphics[scale=0.25]{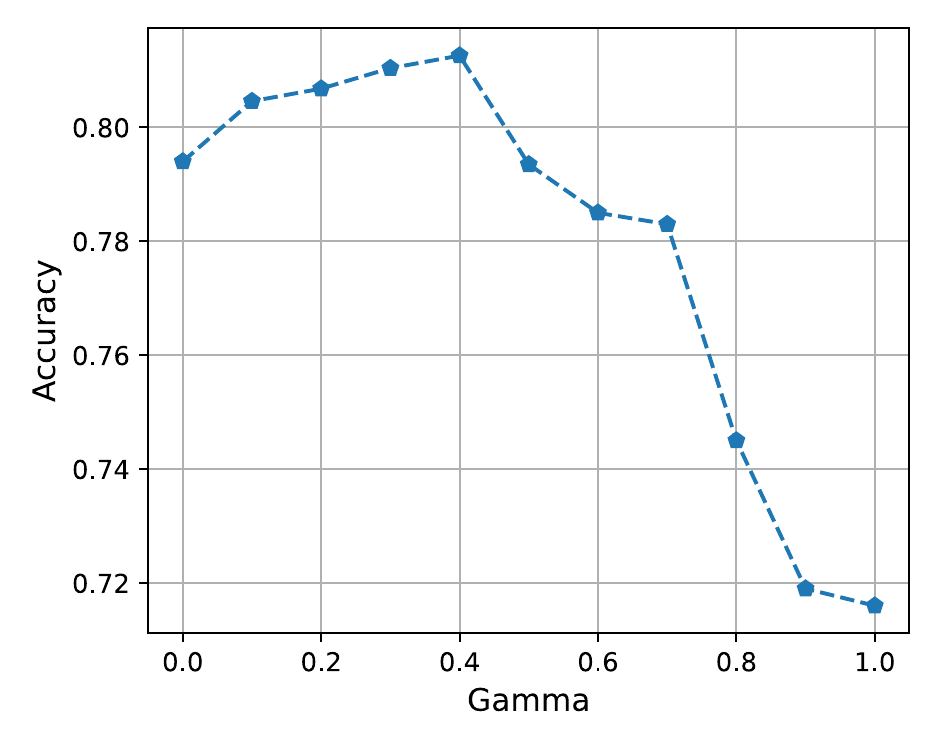}  
\label{fig:overchameleon}
\end{minipage}
}
\subfigure[Citeseer]{   
\begin{minipage}{.23\linewidth}
\centering    
\setlength{\abovecaptionskip}{0.cm}
\includegraphics[scale=0.25]{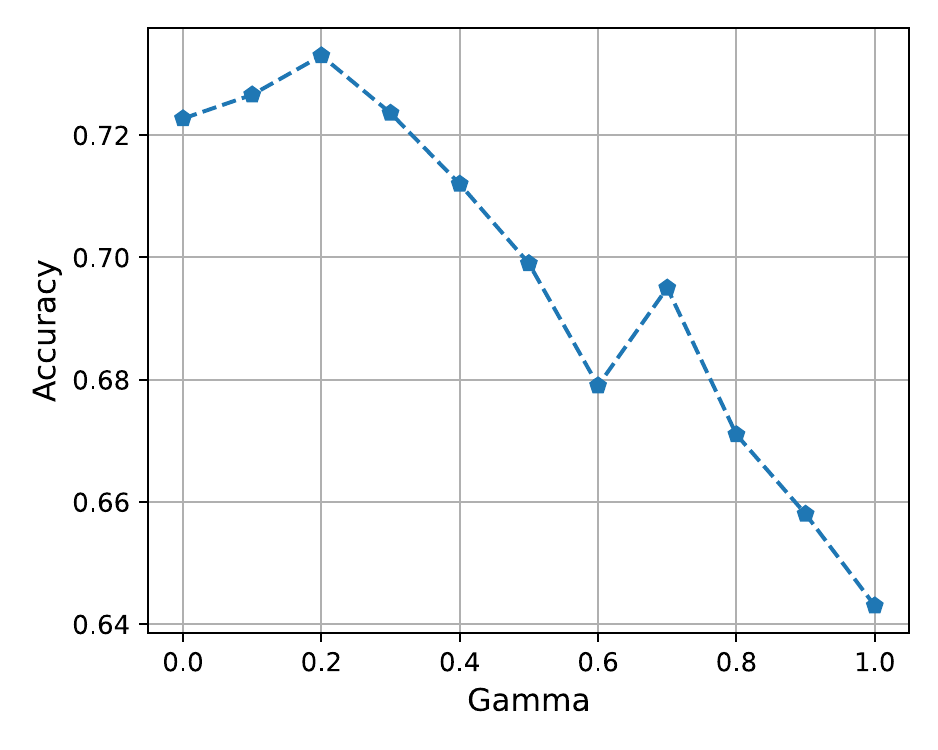}  
\label{fig:overchameleon}
\end{minipage}
}
\subfigure[Actor]{   
\begin{minipage}{.23\linewidth}
\centering    
\includegraphics[scale=0.25]{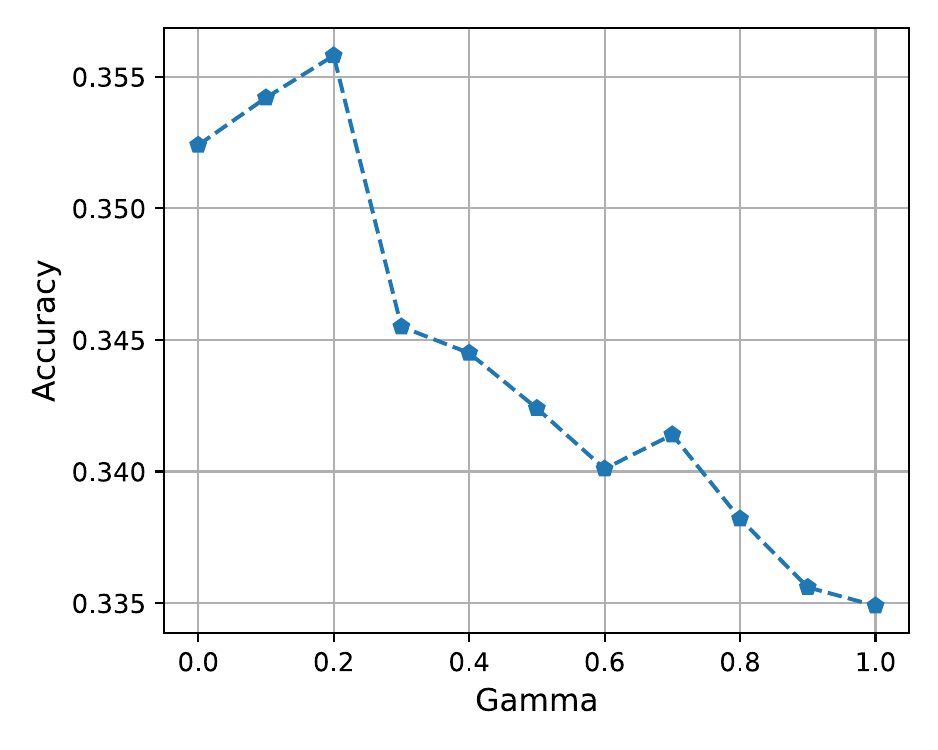}  
\label{fig:overchameleon}
\end{minipage}
}
\caption{Analysis of parameter $\gamma$.}    
\label{fig:r2}    
\end{figure*}

Analysis of parameter $\gamma$ can be found in Figure ~\ref{fig:r2}.

\section{Comparision with other GNN models}
\label{appendix:comparision}
 We compare our model based on DMP with existing methods from the following four perspectives:

(1) Compared with existing discrete GNNs which capture two-hop connectivity (e.g., GCNs, GAT), our model is based on the diffusion equation which generalizes existing discrete GNNs to the continuous case, and the continuous GNNs are fundamentally different from discrete GNNs. The diffusion-based models have unique characteristics. For example, from the perspective of continuous graph diffusion, we can decouple the terminal time and the feature propagation steps, making it more flexible and capable of exploiting a very large number of feature propagation steps, so diffusion-based methods are able to address the common plights of graph learning models such as oversmoothing   \citep{grand, dgc}. Moreover, the diffusion equation can be seen as partial differential equations (PDEs) \citep{grand}, thus introducing many schemes to solve the graph diffusion equation such as explicit scheme, implicit scheme, and multi-step scheme, some of which are more stable and converge faster than normal GNNs.

(2) Compared with the first-order diffusion process, HiD-Net utilizes the 2-hop neighbors' information, where the advantages are: First, it considers a larger neighborhood size and captures the local environment around a node, thus even if there are some abnormal features among the 1-hop neighbors, their negative effect can still be alleviated. Second, the 2-hop neighbors provide an additional stronger correlation with labels according to the monophily property, so even if the 1-hop neighbors may be heterophily, HiD-Net can still make better predictions.

    

(3) Compared with the first-order diffusion process which captures high-order connectivity through iterative adjacent message passing, our proposed second-order diffusion equation is still very different from it. Specifically, the learned representation based on the first-order diffusion process after two propagation steps at time $t$ is:
\begin{equation}
\begin{aligned}
\mathbf{X}^{(t+2\Delta t)}&=(2\alpha \Delta t-\alpha(\alpha+\beta)(\Delta t)^2+\alpha\beta(\Delta t)^2\hat{\tilde{\mathbf{A}}}) \mathbf{X}^{(0)} \\&+ \left[ 1+(\alpha + \beta)^2(\Delta t)^2 -2(\alpha+\beta)\Delta t\right] \mathbf{X}^{(t)}\\
&+\left( 2\beta \Delta t - 2 \beta (\alpha+\beta) (\Delta t)^2\right)\hat{\tilde{\mathbf{A}}}\mathbf{X}^{(t)} \\
&+\beta^2 (\Delta t)^2 \hat{\tilde{\mathbf{A}}}^2\mathbf{X}^{(t)}.
\end{aligned}
\label{eq:2pro}
\end{equation}

After one propagation step at time $t$, HiD-Net has:
\begin{equation}
\begin{aligned}
\mathbf{X}^{(t+\Delta t)}=\alpha \Delta t \mathbf{X}^{(0)} + \left[ 1-(\alpha + \beta)\Delta t \right. \\ \left.+ (\beta - \beta \gamma) \Delta t \hat{\tilde{\mathbf{A}}} +\beta \gamma \Delta t\hat{\tilde{\mathbf{A}}}^2 \right] \mathbf{X}^{(t)}.
\end{aligned}
\label{eq:1hid}  
\end{equation}

Here, in order to facilitate the analysis, we use $\rho_0$, $\rho_1$, $\rho_2$ to denote the coefficient of $\mathbf{X}^{(t)}$, $\hat{\tilde{\mathbf{A}}}\mathbf{X}^t$ and $\hat{\tilde{\mathbf{A}}}^2\mathbf{X}^t$ separately. By analyzing the proportions between different coefficients, we can analyze the weights of neighbors of different orders to see how flexible it is to adjust the balance between neighbors of different orders.

From ~\eqref{eq:2pro}, we have:

\begin{equation}
\setlength{\abovedisplayskip}{3pt}
\setlength{\belowdisplayskip}{3pt}
\rho_0:(\rho_1+\rho_2)=(1-\Delta t)^2:\beta\Delta t(2-2\Delta t + \beta \Delta t)   ,
\label{eq:2pro1}
\end{equation}
\begin{equation}
 \rho_1:\rho_2=2(1-\Delta t):\beta \Delta t .  
 \label{eq:2pro2}
\end{equation}

From ~\eqref{eq:1hid} , we have:

\begin{equation}
\setlength{\abovedisplayskip}{3pt}
\setlength{\belowdisplayskip}{3pt}
\rho_0:(\rho_1+\rho_2)=(1-\Delta t):\beta \Delta t  ,         
\label{eq:1hid1}
\end{equation}
\begin{equation}
\rho_1:\rho_2=(1-\gamma):\gamma .            
\label{eq:1hid2}
\end{equation}

According to ~\eqref{eq:2pro1}, we can adjust the balance between the node itself and its neighbors by adjusting $\beta$ and $\Delta t$. However, as $\beta$ and $\Delta t$ are fixed to adjust $\rho_0:(\rho_1+\rho_2)$, we can't adjust $\rho_1:\rho_2$. That is to say, the proportion of first-order to second-order neighbors is fixed, so the first-order diffusion process can't assign an appropriate proportion for first-order to second-order neighbors.

According to ~\eqref{eq:1hid1}, we can change $\beta$ and $\Delta t$ to adjust the balance between the node itself and its neighbors. According to ~\eqref{eq:1hid2}, we can adjust the balance between first-order neighbors and second-order neighbors by $\gamma$, thus breaking the limitation of the first-order diffusion process.

(4) Time and memory complexity comparison

The propagation step of the first-order and the second-order diffusion have the same memory complexity and time complexity. As $\hat{\tilde{\mathbf{A}}}^2$ can be precomputed, it takes $\mathcal{O}(n^2\zeta)$ to calculate $\hat{\tilde{\mathbf{A}}}^2\mathbf{X}$, where $\mathbf{X}\in\mathbb{R}^{n \times \zeta}$, $n$ is the number of the nodes, $\zeta$ is the dimension of the feature vector. It takes $\mathcal{O}(n^2\zeta)$ to calculate $\hat{\tilde{\mathbf{A}}}\mathbf{X}$, and $\mathcal{O}(1)$ to calculate $\alpha \Delta t \mathbf{X}^{(0)}$ and $(1-(\alpha + \beta)\Delta t)\mathbf{X}^{(t)}$. So the time complexity of the propagation step of HiD-Net is  $\mathcal{O}(n^2\zeta)$ and the first-order diffusion is also $\mathcal{O}(n^2\zeta)$. The memory cost of the first-order diffusion and the second-order diffusion propagation step is $\mathcal{O}(n\zeta)$.

\end{document}